\documentclass[twocolumn]{aastex7}
\usepackage{docmute}
\usepackage{amssymb,amsmath}
\usepackage{physics}
\usepackage{bm}
\usepackage{graphicx}
\usepackage{color}
\usepackage{array}
\usepackage{epsfig}
\usepackage{url}
\usepackage{CJK}
\usepackage{natbib}
\usepackage{threeparttable}
\usepackage{wrapfig}
\usepackage{here}
\usepackage{float}
\usepackage{longtable}
\usepackage{subcaption}
\usepackage{multirow}
\usepackage{xspace}
\usepackage{placeins}
\usepackage{appendix}
\usepackage{booktabs}

\usepackage[T1]{fontenc}

\usepackage[utf8]{inputenc}
\usepackage{cleveref}

\newcommand{\JA}[1]{\begin{CJK*}{UTF8}{ipxm}#1\end{CJK*}}
\newcommand{\CN}[1]{\begin{CJK*}{UTF8}{gbsn}#1\end{CJK*}}
\newcommand{\KR}[1]{\begin{CJK*}{UTF8}{mj}#1\end{CJK*}}

\begin{document}
\begin{CJK*}{UTF8}{ipxm}
\title{\textit{Cosmic Himalayas} in \textit{CROCODILE}: Probing the Extreme Quasar Overdensities by Count-in-Cells analysis and Nearest Neighbor Distribution}

\author[0009-0005-1273-0742]{Yuto Kuwayama (\JA{桑山~裕斗})}
\affiliation{Theoretical Astrophysics, Department of Earth and Space Science, The University of Osaka, 1-1 Machikaneyama, Toyonaka, Osaka 560-0043, Japan}
\email{kuwayama@astro-osaka.jp}

\author[0000-0002-2725-302X]{Yongming Liang (\CN{梁~永明})}
\affiliation{National Astronomical Observatory of Japan, 2-21-1 Osawa, Mitaka, Tokyo 181-8588, Japan}
\email{ymliang@icrr.u-tokyo.ac.jp}

\author[0000-0001-7457-8487]{Kentaro Nagamine (\JA{長峯~健太郎})}
\affiliation{Theoretical Astrophysics, Department of Earth and Space Science, The University of Osaka, 1-1 Machikaneyama, Toyonaka, Osaka 560-0043, Japan}
\affiliation{Theoretical Joint Research, Forefront Research Center, Graduate School of Science, The University of Osaka, Toyonaka, Osaka 560-0043, Japan}
\affiliation{Kavli IPMU (WPI), UTIAS, The University of Tokyo, 5-1-5 Kashiwanoha, Kashiwa, Chiba 277-8583, Japan}
\affiliation{Department of Physics \& Astronomy, University of Nevada, Las Vegas, 4505 S. Maryland Pkwy, Las Vegas, NV 89154-4002, USA}
\affiliation{Nevada Center for Astrophysics, University of Nevada, Las Vegas, 4505 S. Maryland Pkwy, Las Vegas, NV 89154-4002, USA}
\email{kn@astro-osaka.jp}

\author[0000-0002-5712-6865]{Yuri Oku (\JA{奥~裕理})}
\affiliation{Center for Cosmology and Computational Astrophysics, the Institute for Advanced Study in Physics,  Zhejiang University, China} 
\email{oku@astro-osaka.jp}

\author[0009-0007-7716-1290]{Daisuke Nishihama (\JA{西濱~大将})}
\affiliation{Theoretical Astrophysics, Department of Earth and Space Science, The University of Osaka, 1-1 Machikaneyama, Toyonaka, Osaka 560-0043, Japan}
\email{nishihama@astro-osaka.jp}

\author[0000-0003-3467-6079]{Daisuke Toyouchi (\JA{豊内~大輔})}
\affiliation{Theoretical Astrophysics, Department of Earth and Space Science, The University of Osaka, 1-1 Machikaneyama, Toyonaka, Osaka 560-0043, Japan}
\email{toyouchi@astro-osaka.jp}
 
\author[0000-0002-5045-6052]{Keita Fukushima (\JA{福島~啓太})}
\affiliation{Institute for Data Innovation in Science, Seoul National University, Seoul 08826, Korea}
\email{fukushima@snu.ac.kr}

\author[0000-0002-1319-3433]{Hidenobu Yajima (\JA{矢島~秀伸})}
\affiliation{Center for Computational Sciences, University of Tsukuba, 1-1-1 Tennodai, Tsukuba, Ibaraki 305-8577, Japan}
\email{yajima@ccs.tsukuba.ac.jp}

\author[0000-0002-7464-7857]{Hyunbae Park (\KR{박~현배})}
\affiliation{Center for Computational Sciences, University of Tsukuba, 1-1-1 Tennodai, Tsukuba, Ibaraki 305-8577, Japan}
\email{hyunbae.park@gmail.com}

\author[0000-0002-1049-6658]{Masami Ouchi (\JA{大内~正己})}
\affiliation{Institute for Cosmic Ray Research, The University of Tokyo, 5-1-5 Kashiwanoha, Kashiwa, Chiba 277-8582, Japan}
\affiliation{National Astronomical Observatory of Japan, 2-21-1 Osawa, Mitaka, Tokyo 181-8588, Japan}
\affiliation{Department of Astronomical Science, SOKENDAI (The Graduate University for Advanced Studies), Mitaka, Tokyo 181-8588, Japan}
\affiliation{Kavli IPMU (WPI), UTIAS, The University of Tokyo, 5-1-5 Kashiwanoha, Kashiwa, Chiba 277-8583, Japan}
\email{ouchims@icrr.u-tokyo.ac.jp}

\correspondingauthor{Yuto Kuwayama (桑山~裕斗)}
\email{kuwayama@astro-osaka.jp}


\begin{abstract}
The recently reported \textit{Cosmic Himalayas} (CH) -- an extreme quasar overdensity at $z\sim2$-- poses an apparent challenge to the $\Lambda$CDM framework, with a reported significance of $\delta = 16.9\sigma$ under Gaussian assumptions. Such an event appears improbably rare, with a formal probability of $P \sim 10^{-68}$. In this work, we investigate whether CH-like structures can naturally arise in cosmological hydrodynamic simulations. Using the \textit{CROCODILE} simulation, which self-consistently models galaxy--black hole coevolution, we examine quasar clustering through two complementary approaches: the count-in-cells (CIC) statistic, which probes large-scale overdensities, and the nearest-neighbor distribution (NND), sensitive to small-scale environments. CIC analysis reveals that the underlying distribution is heavy-tailed and non-Gaussian, and that conventional Gaussian-based evaluation substantially overestimates the significance of extreme events. When modeled with an asymmetric generalized normal distribution (AGND), the inferred rarity of the CH is substantially reduced and reconciled with standard $\Lambda$CDM; for instance, regions appearing as $12 \sigma_{\rm Gauss}$ outliers under Gaussian assumptions ($P\sim10^{-33}$) are found to occur in AGND regime with a probability of $P\sim10^{-4}$. NND analysis further demonstrates that extreme overdense regions within the simulation can naturally sustain two-point correlation function values similar to those observed in the CH ($r^{\rm eff}_0 \simeq 30 \, h^{-1} {\rm Mpc}$), suggesting that the strong clustering stems from sample selection biases and local environmental variations. These two analyses conclusively highlight the importance of adopting non-Gaussian statistics when quantifying extreme overdensities of quasars and establish that the CH is not an anomaly, but a natural outcome of structure formation in the $\Lambda$CDM universe. 
\end{abstract}

\section{Introduction}
\par Understanding the formation of large-scale structures in the early universe is a central goal of modern cosmology \citep{scoville_cosmic_2007,toni_cosmos-web_2025,croft_large-scale_2025,richarte_quasar_2025}. 
Within the Lambda Cold Dark Matter ($\Lambda$CDM) paradigm, structure grows hierarchically: small-scale fluctuations collapse first and subsequently merge to form larger systems. 
One of the most striking manifestations of this process is the emergence of protoclusters at high redshift, the progenitors of today's galaxy clusters \citep{chiang_ancient_2013, overzier_realm_2016}. 
These overdense regions provide unique laboratories for studying the interplay between galaxy assembly, black hole growth, and feedback processes under extreme conditions. 
Quasar clustering offers a particularly powerful tracer of such environments, as luminous quasars are expected to reside in the most massive halos and thus mark the densest peaks in the cosmic web \citep{garcia-vergara_strong_2017, mignoli_web_2020}. 

\par A remarkable recent discovery is the ``Cosmic Himalayas'' (CH), an exceptionally overdense region of quasars identified by \citet{liang_cosmic_2025}. 
The CH has been interpreted as a protocluster-scale structure \citep{hatch_why_2014,garcia-vergara_alma_2022,wylezalek_first_2022}, but the reported quasar overdensity of $\delta = 16.9\sigma$ presents a striking challenge. 
Under Gaussian assumptions, such an extreme event corresponds to a probability of $P \sim 10^{-68}$, far too rare to be expected in a standard $\Lambda$CDM universe. 
This raises a central question: does the CH represent a genuine anomaly that challenges the cosmological model, does it stem from our limited understanding of the complex astrophysical processes governing quasar formation, or does the extraordinary significance arise from limitations in the statistical methods used to quantify overdensity? 
Resolving this tension is essential for interpreting not only the CH but also the broader role of extreme overdensities in cosmic structure formation. 

\par In this context, cosmological simulations have become indispensable tools. They allow us to investigate structure formation beyond simple analytic approximations, specifically targeting the extreme environments of protoclusters and quasar host regions (e.g., \citealt{Trebitsch2021}).
These simulations enable detailed studies of the co-evolution of galaxies, black holes, and the intergalactic medium in overdense regions. At redshifts around $z \sim 2$, the epoch of peak cosmic star formation and quasar activity, characterizing high-density regions via simulations provides crucial insight into the earliest stages of cluster assembly \citep[e.g.][]{Lim2021, Remus2023,yajima_forever22_2022, Andrews2025}.
Simulation-based studies reveal a diversity of protocluster evolutionary states and highlight challenges in reproducing the high star-formation rates observed in dense regions \citep{overzier_realm_2016}. 
Such efforts signify the importance of simulations for interpreting extreme overdensities within the $\Lambda$CDM framework. 

\par To address this  specifically, we analyze the cosmological hydrodynamic simulation \textit{CROCODILE} \citep{oku_osaka_2024}, which self-consistently models the growth of galaxies and supermassive black holes. 
This framework allows us to search for CH analogs and to test whether such extreme quasar overdensities can arise naturally within $\Lambda$CDM. 
Identifying these analogs in simulations is also a crucial first step toward exploring their observed correlations with H\,\textsc{i} gas and Ly$\alpha$ emitters, providing a pathway to investigate quasar feedback on the surrounding intergalactic medium. 

\par Our analysis employs two complementary statistical tools. 
The count-in-cells (CIC) method provides a one-point statistic that evaluates the distribution of quasar counts across fixed volumes. 
By constructing the probability distribution $P_{\rm CIC}(N; V)$ of counts within cells of size $V$, the CIC directly measures deviations from a random (Poisson) distribution and thus probes the statistical rarity of overdense regions \citep[e.g.][]{ueda_counts--cells_1996,yang_galaxy_2011,shin_new_2017}. 
Since CIC was the method used by \citet{liang_cosmic_2025} to evaluate the CH, it provides a direct point of comparison. 
Complementing the CIC, the nearest-neighbor distribution (NND) quantifies the distances to the closest quasar pairs and is particularly sensitive to compact structures and dense environments. NND statistics provide a powerful diagnostic of small-scale clustering and local environment \citep{ryden_statistical_1984}, and their use has expanded in both cosmology and related fields \citep{BanerjeeAbel2021,ouellette_cosmological_2025, zhou_high-energy_2025}.

\par Throughout this work, we assume the cosmological parameters from 
\citet{planck_collaboration_planck_2020}: $(\Omega_m, \Omega_b, \Omega_\Lambda, h, n_s, 10^9 A_s) =  (0.3099, 0.0488911, 0.6901, 0.67742, 0.96822, 2.1064)$.
The structure of this paper is as follows: 
Section~\ref{sec:simulation} outlines the simulation setup, 
Section~\ref{sec:analysis} details the methodology for the CIC and NND analyses and presents the results,
Section~\ref{sec:discussion} discusses the emergence of extreme overdense regions of quasars in the context of large-scale observations, 
and Section~\ref{sec:summary} summarizes our conclusions.

\begin{table*}[btph]
\caption{Parameter-set of the simulation box of each simulation.}
    \centering
    \begin{tabular}{ccc} \hline \hline
        Parameters & Unit & 
        \begin{tabular}{c}
            \textsc{L200N1024 BoostFactor50 No Eddington Limit} \\ 
            (\texttt{BF50NEL})
        \end{tabular} 
        \\ \hline
        Boxsize &  $(h^{-1}{\rm cMpc})^3$ & $200^3$ \\  
        Target Redshift of this work & - & 2.2 \\
        Softening length & $h^{-1}\, {\rm ckpc}$ & 6.76  \\
        Maximum softening length & $h^{-1}\, {\rm pkpc}$ & 1.00 \\        
        DM mass &  $h^{-1}M_{\odot}$ & $5.39 \times 10^8$ \\
        Gas mass &  $h^{-1}M_\odot$ & $1.01 \times 10^8$ \\
        Seed BH mass &  $h^{-1}M_\odot$ & $1 \times 10^5$ \\
        $C_{\rm visc}$ & - & $200\pi$ \\
        Boost Factor & - & 50  \\ 
        Calculation of BH accretion & - & 
        \begin{tabular}{c}
            Angular-momentum-limited Bondi accretion \\
            w/o the Eddington accretion limit  
        \end{tabular} \\ \hline
    \end{tabular}
    
    \label{tab:parameter-set}
\end{table*}

\section{Cosmological Hydrodynamic Simulation \textit{CROCODILE}} 
\label{sec:simulation}

\par Our analysis is based on the \textit{CROCODILE}\footnote{CROCODILE Project: \url{https://sites.google.com/view/crocodilesimulation/home}} v2 suite of cosmological hydrodynamic simulations, performed with the \textsc{gadget4-osaka} smoothed particle hydrodynamics (SPH) code \citep{romano_dust_2022,romano_co-evolution_2022,oku_osaka_2024},  
which is built upon the publicly available \textsc{gadget-4} code \citep{springel_simulating_2021} \footnote{\textsc{gadget-4}: \url{https://wwwmpa.mpa-garching.mpg.de/gadget4/}}.  Detailed descriptions of the gravity and hydrodynamics solvers can be found in \citet{springel_simulating_2021} and \citet{oku_osaka_2024}. 
\par The initial condition is generated using MUSIC2-MONOFONIC\footnote{MUSIC2-MONOFONIC: \url{https://bitbucket.org/ohahn/monofonic/src/master/}} \citep{hahn_higher_2021, michaux_accurate_2021}. To suppress the impact of cosmic variance, we employed the phase-fixing technique \citep{angulo_cosmological_2016}. The particles were initialized at $z_{\rm ini} = 39$ using third-order Lagrangian perturbation theory (3LPT). The transfer function at $z_{\rm ini}$ was obtained by back-scaling the transfer function from a reference redshift of $z_{\rm ref} \sim 2$, computed by CLASS\footnote{CLASS: \url{http://class-code.net/}} \citep{blas_cosmic_2011}. 

\par The simulation volume is a periodic cubic box of side length $L_{\rm box}=200 \, h^{-1} \, {\rm cMpc}$. We employ $1024^3$ dark matter particles and an initially equal number of gas particles. This configuration corresponds to a dark matter particle mass of $m_{\rm DM} = 5.39 \times 10^8 \, h^{-1} \, M_\odot$ and initial gas particle mass of $m_{\rm gas} = 1.01 \times 10^8 \, h^{-1} \, M_\odot$. The gravitational softening length is set to $\epsilon_{\rm grav} = 6.76 \, h^{-1} \, {\rm ckpc}$, but limited to physical $1.00 \, h^{-1} \, {\rm pkpc}$ at all times. 

\par The simulations incorporate sub-grid prescriptions for key baryonic processes, including star formation, supernova (SN) feedback \citep{oku_osaka_2022}, and active galactic nucleus (AGN) feedback \citep{oku_osaka_2024}. 
Chemical enrichment is modeled with the CELib library \citep{saitoh_chemical_2017}, which self-consistently tracks the enrichment of multiple elements. 

\par Radiative cooling and heating are computed using the \textsc{grackle}\footnote{We use version \textsc{grackle}-3.2.1, see \url{https://grackle.readthedocs.io/en/grackle-3.2.1/.}} chemistry and cooling library \citep{smith_grackle_2017}. 
\textsc{grackle} solves a non-equilibrium primordial chemical network including hydrogen, deuterium, helium species, and molecular hydrogen (H$_2$) and HD. 
Metal-line cooling is incorporated via precomputed tables from the photoionization code \textsc{Cloudy} \citep{ferland_2013_2013}, which account for photoionization and photoheating by the cosmic ultraviolet background (UVB), with self-shielding corrections applied at high density. 
We adopt the spatially uniform, redshift-dependent UVB model of \citet{haardt_radiative_2012}, which is activated at $z=8$.

\subsection{BH accretion and AGN feedback model } \label{sec:BHaccretion}
\par First, we describe the seeding prescription for black holes (BHs).
We employ a method based on the on-the-fly halo identification using the Friends-of-Friends (FoF) algorithm. A seed BH with an initial mass of $M_{\rm BH} = 10^5 \, h^{-1} \, M_\odot$ is placed at the position of the minimum gravitational potential within a halo. Seeds are introduced only in haloes that exceed a threshold mass of $M_{\rm FoF,seed} = 10^{10} \, h^{-1} \, M_\odot$.
\par Once seeded, the BH accretion rate, $\dot{M}_{\rm acc}$, is calculated using an angular-momentum-limited Bondi prescription with a resolution-dependent boost factor:
\begin{align}
\label{eq:accretion}
    \dot{M}_{\rm acc} = \alpha \, C \, \dot{M}_{\rm Bondi} \, ,
\end{align}
where $\dot{M}_{\rm Bondi}$ is the Bondi accretion rate, and $\alpha = 50$ is the boost factor introduced to compensate for the limited spatial resolution and unresolved small-scale physics such as chaotic cold accretion \citep{gaspari_chaotic_2013}. 
The value $\alpha=50$ is chosen to be consistent with previous implementations in the \textsc{gadget4-osaka} simulations, calibrated to reproduce observed BH mass functions and scaling relations. 
The dimensionless coefficient $C$ accounts for angular-momentum support and viscous effects in the accretion flow, and is defined as \citep{rosas-guevara_impact_2015,oku_osaka_2024}
\begin{align}
\label{C_params}
    C = \min \qty[ C_{\rm visc}^{-1} \qty( \frac{c_s}{v_\phi} )^3 , 1 ] \, ,
\end{align}
where $c_s$ is the sound speed of the ambient gas, $v_\phi$ is the rotational velocity of the gas around the BH, and $C_{\rm visc} = 200\pi$ is a calibration parameter. 
The underlying Bondi accretion rate is given by
\begin{align}
\label{eq:Bondi_accretion}
    \dot{M}_{\rm Bondi} = \frac{4 \pi G^2 M_{\rm BH}^2 \rho}{c_s^3}
\end{align}
where $\rho$ is the local gas density.

\par In the fiducial {\it CROCODILE} runs, BH growth is capped at the Eddington limit. 
In this work, however, we adopt a model without this restriction (``NEL''; No-Eddington-Limit), allowing for brief episodes of super-Eddington accretion, a scenario supported by recent observational detections of super-Eddington AGNs \citep[e.g.,][]{suh_super-eddington-accreting_2025}. 
This choice is motivated by the shortcomings of the Eddington-limited model, which tends to overproduce AGN by failing to sufficiently quench low-mass BHs. 
By enabling short bursts of rapid growth, the NEL model generates more energetic feedback events that effectively suppress accretion in less massive BHs, while simultaneously allowing the most massive BHs to reach the high accretion rates required to reproduce the observed Eddington ratio distribution and luminous quasar population at high redshift. 
As shown in Sec.~\ref{sec:results:BH}, this approach yields a BH population in much better agreement with observational constraints.
  
\par Feedback from BH accretion is implemented as purely thermal AGN feedback, following the EAGLE scheme as adopted in {\it CROCODILE} \citep{schaye_eagle_2015,crain_eagle_2015,oku_osaka_2024}. 
A fixed fraction of the accreted rest-mass energy is coupled to the surrounding gas:
\begin{align}
    \dot{E}_{\rm AGN} = \epsilon_f \, \epsilon_r \, \dot{M}_{\rm acc} c^2 \, ,
\end{align}
where $\epsilon_r = 0.1$ is the radiative efficiency and $\epsilon_f = 0.15$ is the feedback efficiency \citep[see Table~1 of][]{oku_osaka_2024}. 
The energy is injected stochastically into neighboring gas particles as thermal energy, a method designed to avoid numerical overcooling. 
This prescription has been shown to reproduce observed galaxy–BH scaling relations and provides a physically motivated mechanism for regulating BH growth and quasar feedback in cosmological simulations.

\par We also updated the treatment of the sound speed $c_s$ in the accretion model between version~1 (``V1''; \citealt{oku_osaka_2024}) and version~2 (``V2''; Nishihama et al., in prep.) 
In V1, we calculated $c_s$ using only the temperature derived from the effective equation of state (EoS) employed in the EAGLE simulation \citep{schaye_eagle_2015, crain_eagle_2015}: 
\begin{align}
    c_{s,\rm EoS} = \sqrt{\gamma \frac{k_B}{\mu_a m_p} T_{\rm EoS}}
\end{align}
where $\gamma=5/3$ is the adiabatic index, $k_{\rm B}$ is the Boltzmann constant, $\mu = 1.2285$ is the mean molecular weight of primordial atomic gas, and $m_p$ is the proton mass. 
The effective temperature is given by $T_{\rm EoS} = 8 \times 10^3\,{\rm K}\,(n_{\rm H} / 0.1\,{\rm cm^{-3}})^{1/3}$. This formulation in V1 was based on the simplified assumption that BHs always accrete from the star-forming ISM governed by this polytropic EoS. 

\par In V2, we modified this implementation to align with the standard strategy of the EAGLE simulation, where the effective EoS serves as a lower limit (i.e., a temperature floor) for the gas temperature. Specifically, V2 adopts the larger of the EoS temperature and the actual thermodynamic gas temperature $T_{\rm gas}$: 
\begin{align}
    c_{s} &= \sqrt{\gamma \frac{k_B}{\mu_a m_p} T_{\rm v2}} \, , \\
    T_{\rm v2} &= \max \left( T_{\rm EoS}, \, T_{\rm gas} \right) \, ,
\end{align}
where $T_{\rm gas}$ denotes the thermodynamic temperature of the local gas surrounding the BH. This prescription effectively prevents unphysically high Bondi accretion rates in cold, dense gas while accounting for the thermal state of gas hotter than the effective EoS.
Consequently, the Bondi rate is more strongly suppressed in such regions, resulting in systematically weaker AGN feedback than in V1. 
A complete list of the adopted simulation parameters is provided in Table~\ref{tab:parameter-set}. 


\par In practice, this refinement defines the V2 accretion model. 
For the present study, we employ the \texttt{BF50NEL} run, which combines the V2 prescription with the allowance for super-Eddington accretion. 
Together, these features are crucial for reproducing the observed abundance of luminous quasars at high redshift (Figures~\ref{fig:lambda_Edd_distribution}--\ref{fig:BH_mass_function}), making our simulation a robust framework for investigating extreme structures such as the Cosmic Himalayas.

\section{Analysis Methods and Results} \label{sec:analysis}

\par Before detailing the black hole and luminosity selections, we first outline the overall strategy for ensuring a consistent comparison between the simulation and the observations.

\begin{table*}
    \centering
    \caption{Parameter sets and the results of CIC and NND analyses}
    \label{tab:selection_criteria}
    \setlength{\tabcolsep}{2.5pt}
    \begin{tabular}{ccccccc} \hline \hline
                & Simulation & Selection Criteria & $L_{\rm cell} [h^{-1} \, {\rm cMpc}]$ & $\bar{N}_{\rm CIC}$  & $N^{\rm max}_{\rm CIC}$  & $r^{\rm eff}_0 \, [h^{-1} \, {\rm cMpc}]$ \\ \hline
         Case~A &  \textit{CROCODILE} & 
         \begin{tabular}{c}
         $8.0 \leq \log_{10} (M_{\rm BH}/M_\odot) \leq 9.5$, \\
         $\log_{10} (L_{\rm bol}/{\rm erg \, s^{-1}}) >45.47$  \\ 
         w/ Type II reduction \citep{merloni_incidence_2014}
         \end{tabular}
         & 30 & 0.25 & 6 & 30.9\\ \hline
         Case~B &  \textit{CROCODILE} & 
         \begin{tabular}{c}
         $7.1 \leq \log_{10} (M_{\rm BH}/M_\odot) \leq 9.9$, \\
         $\log_{10} (L_{\rm bol}/{\rm erg \, s^{-1}}) >45$ \\ 
         w/ Type II reduction \citep{merloni_incidence_2014}
         \end{tabular}
         & 30 & 1.5 & 17 & 13.7 \\ \hline
        Case~C  &  \textit{Uchuu} (Appendix \ref{appendix:Uchuu}) & 
         \begin{tabular}{c}
         $\log_{10} (M_{200c}/M_\odot) >11.5$ \\ 
         w/ AGN selection \citep{rembold_sdss-iv_2023} 
         \end{tabular}
         & 30  & 0.36 & 9 & -- \\ \hline
    \end{tabular}
\end{table*}

\subsection{Normalization of quasar number density and methodological overview}

\par The observed mean quasar number per $L_{\rm cell} \simeq 30\,h^{-1}\,{\rm Mpc}$ cell, $\bar{N}_{\rm SDSS} = 0.36$, is derived from the SDSS, which covers a comoving Gpc-scale volume and therefore provides a robust estimate of the cosmic mean density of luminous quasars at $z \sim 2$ \citep{liang_cosmic_2025}. 
To enable a meaningful comparison, we apply two quasar selection criteria for our simulation samples. These selection criteria significantly alter cosmic quasar number densities, and varying selections allow us to quantify the sensitivity of the results to the sample selection and assess the impact of different levels of bias.

The fiducial selection criterion (Case~A) gives us a similar number density as the SDSS samples ($\bar{N}_{\rm QSO} \sim 0.25$ in $L_{\rm cell} = 30 \, h^{-1} \, {\rm Mpc}$ cells), as we describe in Section \ref{sec:BH_selection}. On the other hand, the Case~B criterion covers a slightly wider BH mass range than the SDSS samples, resulting in a higher number density than in the SDSS sample ( $\bar{N}_{\rm QSO} \sim 1.5$ in $L_{\rm cell} = 30 \, h^{-1} \, {\rm Mpc}$ cells). 
\par First, we perform the CIC analysis. This characterizes the global distribution and statistical rarity of extreme overdensities such as the CH. For the CIC analysis, a quasar selection criterion sets the mean of the probability distribution $P_{\rm CIC}(N_{\rm QSO}; L_{\rm cell})$ when fitting analytic models such as the asymmetric generalized normal distribution (AGND), while still allowing the higher moments (variance, skewness, and tail shape) to reflect the intrinsic clustering in the simulation.
\par Second, we perform the NND analysis, which directly compares the small-scale clustering of extreme overdensities. This allows us to measure the clustering properties as a probability distribution for finding the nearest neighbors $P_{\rm NN}(<r)$.

\par Although the finite volume of the {\it CROCODILE} box ($L_{\rm box}=200\,h^{-1}\,{\rm Mpc}$) introduces cosmic variance in the estimate of $\bar{N}_{\rm QSO}$, this does not bias the NND analysis, as the normalization ensures consistency with the observationally determined mean density. 
The two statistical measures, therefore, play complementary roles: the CIC quantifies the global distribution and non-Gaussianity, whereas the NND measures the small-scale clustering of quasar overdensities.  

\par We note that the \textit{CROCODILE} simulations employ phase-fixed initial conditions to suppress cosmic variance. While phase fixing does not guarantee unbiased higher-order $N$-point statistics in the non-linear regime, our analysis relies on relative comparisons within the same realization. The persistence of qualitatively similar CIC behavior in the \textit{Uchuu} simulation (\citealt{ishiyama_uchuu_2021}, Appendix \ref{appendix:Uchuu}) with random-phase initial conditions further supports the robustness of our conclusions.

\par To further validate the robustness of our results against box-size limitations, we performed an independent CIC analysis using the {\it Uchuu} simulation \citep{ishiyama_uchuu_2021}, which spans a much larger comoving volume of $L_{\rm box}=2\,h^{-1}\,{\rm Gpc}$. 
As presented in Appendix~\ref{appendix:Uchuu}, this analysis confirms that the non-Gaussian shape of the CIC distribution persists on Gpc scales. This cross-check demonstrates that the conclusions drawn from our fiducial simulation are not artifacts of limited volume, but reflect generic statistical properties of the quasar population within the $\Lambda$CDM framework.



\subsection{Luminosity Estimation of Black hole}
\label{sec:luminosity_estimation}

We estimate the bolometric luminosities of AGNs ($L_{\rm bol}$) as follows:
\begin{align}
    \label{eq:luminosity-accretionrate}
    L_{\rm bol} &= \epsilon_r \dot{M}_{\rm acc} c^2, 
\end{align}
where $\epsilon_r = 0.1$ is the radiative efficiency in the {\it CROCODILE} feedback model. 

\par To construct the X-ray luminosity function (XLF) of AGN, we also estimate the hard X-ray (2--10 keV) luminosity $L_{\rm X}$ from $L_{\rm bol}$. 
For this conversion, we employ the bolometric correction model of \citet{hopkins_observational_2007}, defined as
\begin{align}
    \label{eq:correction_X-ray_luminosity}
    \log_{10} \qty(\frac{L_{\rm X}}{L_\odot}) = \log_{10} \qty(\frac{L_{\rm bol}}{L_\odot}) & - \log_{10} {\rm BC}\, , \\
    {\rm BC} = 10.83 \qty(\frac{L_{\rm bol}}{10^{10} L_{\odot}})^{0.28} &+ 6.08 \qty(\frac{L_{\rm bol}}{10^{10} L_\odot})^{-0.020}\, ,
\end{align}
where $L_\odot$ is the solar luminosity, and ${\rm BC}$ is the bolometric correction factor from $L_{\rm bol}$ to $L_{\rm X}$.

\subsection{Selection Criteria for Quasars from the Simulation}
\label{sec:BH_selection}

\par In our analysis, luminous AGNs are identified as BHs that satisfy both mass and luminosity thresholds. We adopt two alternative sets of selection criteria, denoted as Case~A and Case~B. 
The conditions are summarized below: 

\begin{align}
    \label{eq:mass_condition_A}
    8.0 \leq \log_{10} \left(  \frac{M_{\rm BH}}{M_\odot} \right ) \leq 9.5 \quad (\mathrm{case~A}) \\
    \label{eq:luminosity_condition_A}
    \log_{10} \left( \frac{L_{\rm bol}}{\rm erg \; s^{-1}} \right )  > 45.47  \quad (\mathrm{case~A}) 
\end{align}

\begin{align}
    \label{eq:mass_condition_B}
    7.1 \leq \log_{10} \left(  \frac{M_{\rm BH}}{M_\odot} \right ) \leq 9.9 
    \quad (\mathrm{case~B})  \\
    \label{eq:luminosity_condition_B}
    \log_{10} \left( \frac{L_{\rm bol}}{{\rm erg\, s^{-1}}} \right ) > 45  \quad (\mathrm{case~B}). 
\end{align}

\par Case~A represents a narrower selection, with a BH mass range consistent with the $1\sigma$ scatter of SDSS quasars at $z\sim2$ \citep{rakshit_spectral_2020} and a luminosity threshold comparable to that of the quasars identified in the CH \citep{liang_cosmic_2025}. 

\par Case~B, by contrast, adopts a broader mass interval consistent with the $3\sigma$ scatter of SDSS quasars together with a luminosity threshold reflecting the $2\sigma$ range of the SDSS quasar population at $z \sim 2$. These dual selections allow us to determine the extent to which varying the sample bias influences the resulting clustering properties. 
Together, these two complementary definitions bracket the observational uncertainties, enabling a systematic evaluation of whether CH analogs can naturally arise within the $\Lambda$CDM framework. 
\par For the luminous AGNs, we select Type I AGNs from the total sample via random selection based on the Type I fraction model from \citet{merloni_incidence_2014}: 

\begin{align}
    \label{eq:Obscured_Fraction}
    f_{\rm Type I} = A - \frac{1}{\pi} \arctan \qty( \frac{l_0 - l_x}{\sigma_x}) 
\end{align}
where $l_x = \log_{10} L_{\rm X}, A = 0.44, l_0 = 43.89, \sigma_x = 0.46$. This model introduces stochasticity. To account for this sample-selection uncertainty, we generated 100 independent realizations of the Type I AGN sample via random sampling and analyzed the results based on the ensemble in each case. Table \ref{tab:selection_criteria} summarizes the selection criteria and the corresponding analysis results for each case.




\subsection{AGN properties in CROCODILE} 
\label{sec:results:BH}

\begin{figure}[t]
    \centering
    \includegraphics[width=\linewidth]{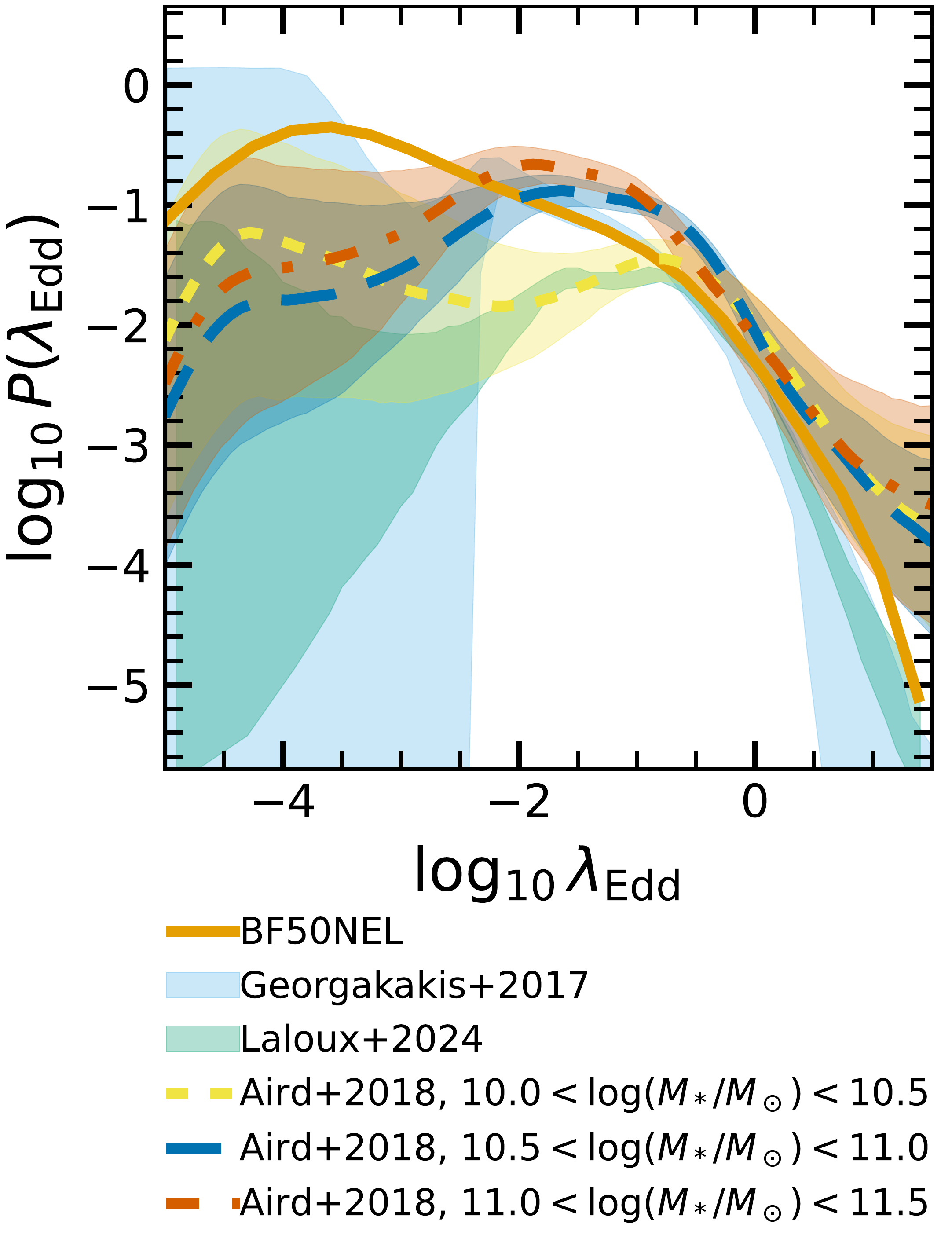}
    \caption{Eddington ratio ($\lambda_{\rm Edd}$) distribution function for BHs in our simulation BF50NEL (solid orange line).  Colored lines and shaded regions indicate observational constraints: sky blue and green shades correspond to \citet{georgakakis_observational_2017} ($z=2.25$) and \citet{laloux_accretion_2024} ($1.5 < z<2.0$), respectively, while yellow, blue, and thick orange dotted lines with shades represent the constraints in stellar mass bins of \citet{aird_x-rays_2018} ($2.0 < z < 2.5$).}
    \label{fig:lambda_Edd_distribution}
\end{figure}

\par To clarify the BH properties in our simulation, we examine the Eddington ratio distribution function (ERDF) (Eddington ratio; $\lambda_{\rm Edd} \equiv \dot{M}_{\rm BH}/\dot{M}_{\rm Edd}$), the X-ray luminosity function (XLF), and the BH mass function (BHMF), and compare them with both observational data and other simulation results. 

\par Figure~\ref{fig:lambda_Edd_distribution} presents the ERDF measured from our \texttt{BF50NEL} run, alongside several observational estimates. The observational constraints are taken from \citet{georgakakis_observational_2017}, \citet{laloux_accretion_2024}, and \citet{aird_x-rays_2018}, corresponding to redshift ranges $z=2.25$ \citep{georgakakis_observational_2017}, $1.5 < z < 2.0$ \citep{laloux_accretion_2024}, and $2.0 < z < 2.5$ \citep{aird_x-rays_2018}, respectively. 
For \citet{aird_x-rays_2018}, we show results in three stellar mass bins: $10.0 < \log_{10}(M_*/M_\odot) < 10.5$, $10.5 < \log_{10}(M_*/M_\odot) < 11.0$, and $11.0 < \log_{10}(M_*/M_\odot) < 11.5$. 
\par Notably, the ERDF of our \texttt{BF50NEL} run (Figure~\ref{fig:lambda_Edd_distribution}) shows excellent agreement with these observational constraints, particularly in the high-Eddington-ratio regime (i.e., $\log_{10}\lambda_{\rm Edd} \gtrsim-2.5$). This consistency indicates that our model successfully reproduces the population of bright, actively accreting BHs that are relevant to the quasar activity discussed in this work.

\begin{figure}[t]
    \centering   
    \includegraphics[width=\linewidth]{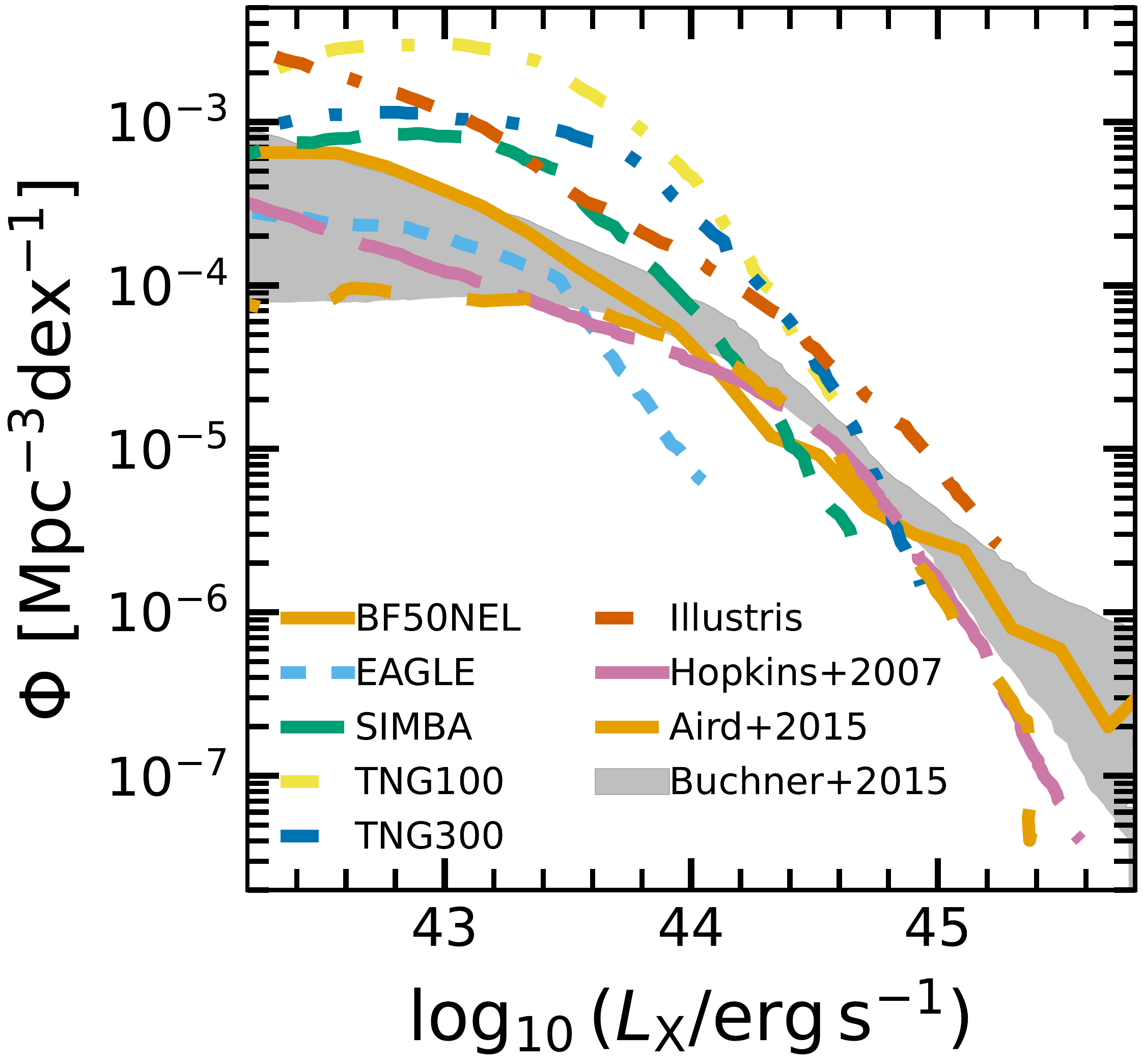}
    \caption{X-ray luminosity function (XLF) from simulations and observations. The orange solid line shows the result from our \texttt{BF50NEL} simulation. Colored dotted curves indicate results from other simulation projects compiled by \citet{habouzit_supermassive_2022}: EAGLE (sky blue; \citealt{schaye_eagle_2015,crain_eagle_2015}), SIMBA (green; \citealt{dave_simba_2019}), TNG100 (yellow) and TNG300 (blue; \citealt{nelson_first_2018}), and Illustris (red; \citealt{vogelsberger_properties_2014}). Observational estimates are shown from \citet{hopkins_observational_2007} (pink), \citet{aird_evolution_2015} (orange dotted line), and \citet{buchner_obscuration-dependent_2015} (gray shade).}
    \label{fig:XLF}
\end{figure}

\par Figure~\ref{fig:XLF} presents the X-ray luminosity function (XLF) from our simulation, compared with both observational data and results from other large cosmological simulations. The comparison simulation results are taken from the analysis of \citet{habouzit_supermassive_2022}:  EAGLE \citep{schaye_eagle_2015,crain_eagle_2015}, SIMBA \citep{dave_simba_2019}, TNG100 and TNG300 \citep{nelson_first_2018}, and Illustris \citep{vogelsberger_properties_2014}. Crucially, we estimate the hard X-ray luminosity (2--10 keV) applying the bolometric correction of \citet{hopkins_observational_2007}, consistent with the methodology of \citet{habouzit_supermassive_2022}.
The observational constraints shown are from \citet{hopkins_observational_2007}, \citet{aird_evolution_2015}, and \citet{buchner_obscuration-dependent_2015}.

\par The XLF of our \texttt{BF50NEL} run (Figure~\ref{fig:XLF}) shows good agreement with the \citet{buchner_obscuration-dependent_2015} results at $\log_{10} (L_X/{\rm erg \, s^{-1}}) \lesssim 44$ and $\log_{10} (L_X/{\rm erg \, s^{-1}}) \gtrsim 44.8$, though it slightly underestimates the number densities in the intermediate range $44 \lesssim \log_{10} (L_X/{\rm erg \, s^{-1}}) \lesssim 44.8$.  

\par
\begin{figure}[t]
    \centering
    \includegraphics[width=\linewidth]{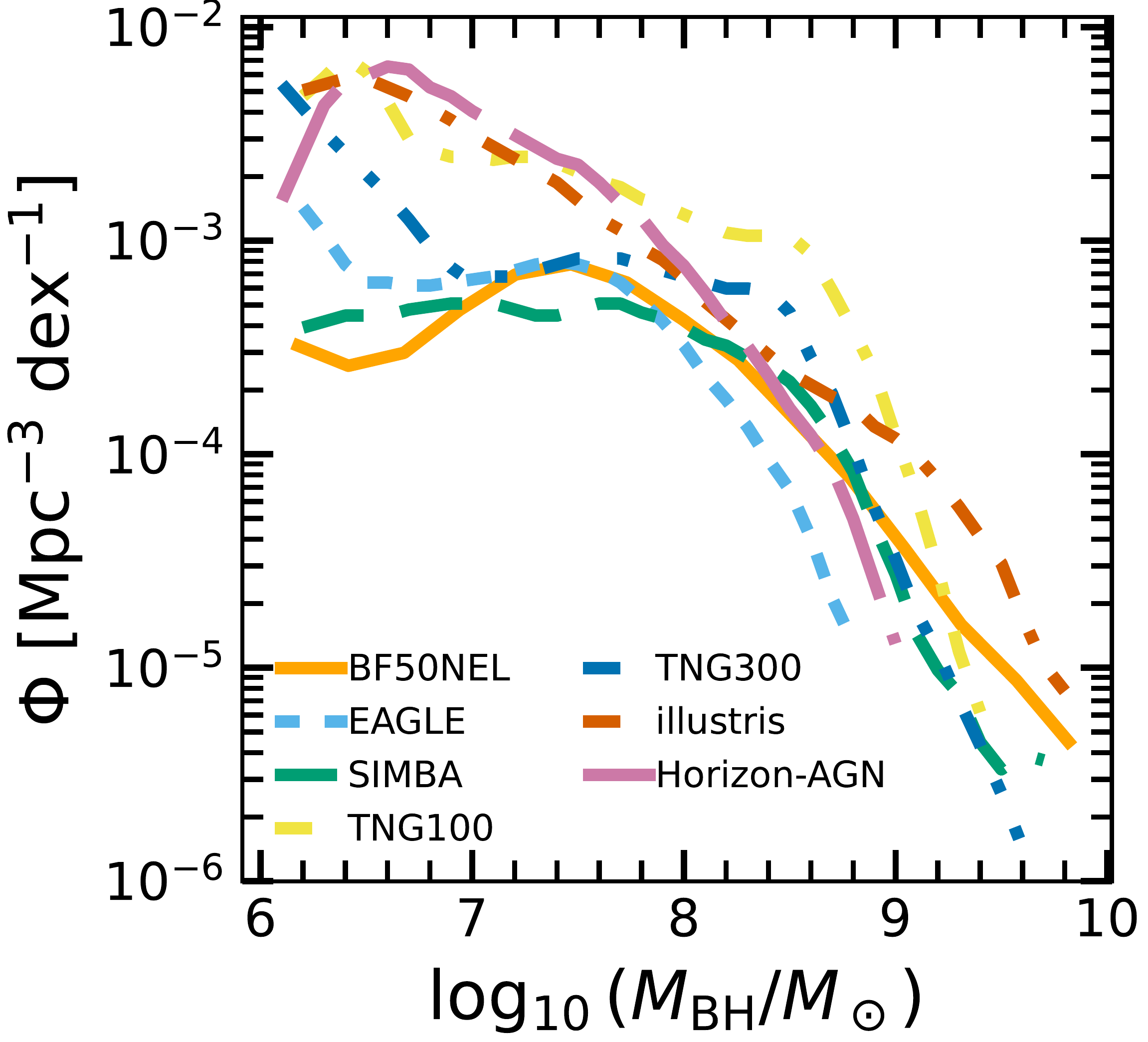}
    \caption{Black hole mass function (BHMF) from simulations. The orange solid line shows the result from our \texttt{BF50NEL} simulation. Other colored curves represent results from major cosmological hydrodynamic simulations compiled by \citet{habouzit_supermassive_2021}: EAGLE (sky blue; \citealt{schaye_eagle_2015,crain_eagle_2015}), SIMBA (green; \citealt{dave_simba_2019}), TNG100 (yellow) and TNG300 (blue; \citealt{nelson_first_2018}), Illustris (red; \citealt{vogelsberger_properties_2014}), and Horizon-AGN (pink; \citealt{dubois_dancing_2014}). }
    \label{fig:BH_mass_function}
\end{figure}

 Figure~\ref{fig:BH_mass_function} shows the BH mass function (BHMF) from our simulation, together with other simulations analyzed by \citet{habouzit_supermassive_2021}: the same set as in the XLF comparison (EAGLE, SIMBA, TNG100, TNG300, and Illustris), along with Horizon-AGN \citep{dubois_dancing_2014}.  

\begin{figure*}[t]
    \centering
    \begin{minipage}[t]{0.437\linewidth}
        \includegraphics[width=\linewidth]{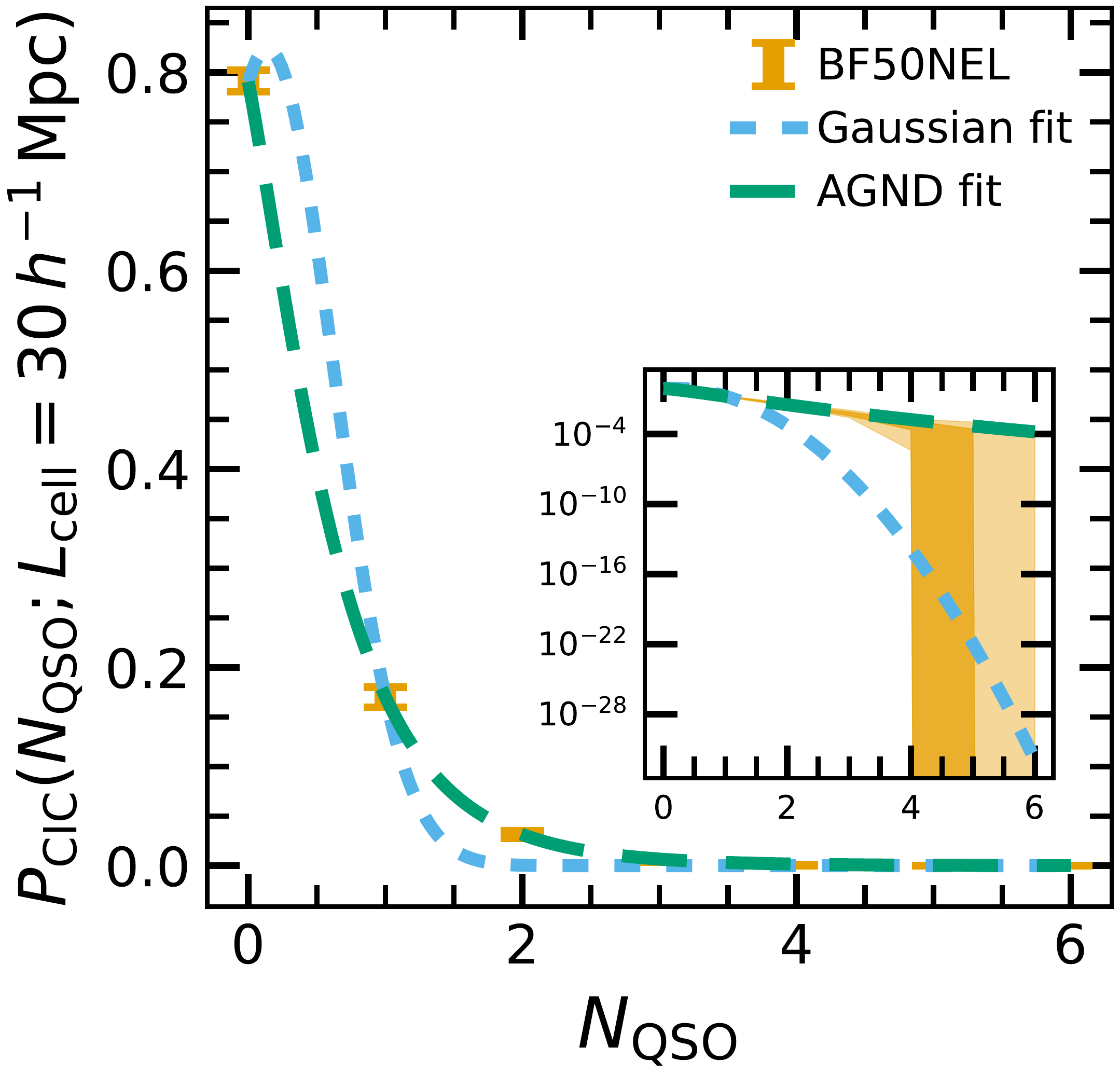}
        \subcaption{CIC analysis for Case~A. 
        $N_{\rm QSO}^{\rm max} = 6\; ( \delta_{\rm QSO}^{\rm max}\sim23)$.}
    \end{minipage}
    \hspace{0.05\linewidth}
    \begin{minipage}[t]{0.45\linewidth}
        \includegraphics[width=\linewidth]{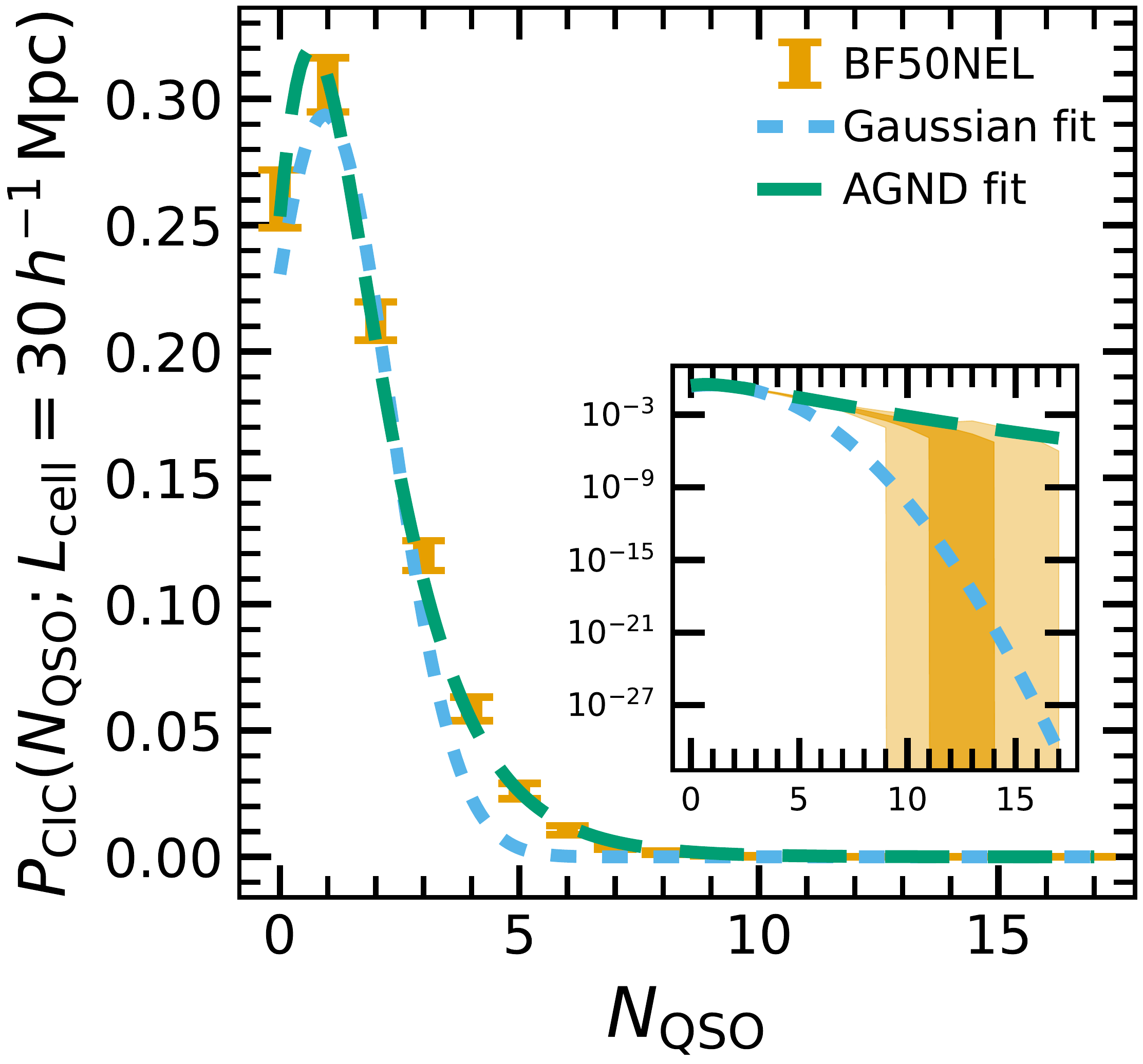}
        \subcaption{CIC analysis for Case~B. 
        $N_{\rm QSO}^{\rm max} = 17 \; (\delta_{\rm QSO}^{\rm max} \sim 10)$.}
    \end{minipage}
    
    \caption{Count-in-cells (CIC) probability distribution of quasars for  $L_{\rm cell}=30\,h^{-1}\,{\rm Mpc}$. The orange error bars show the CIC measured from our simulation (\texttt{BF50NEL}), which corresponds to the scatter obtained from the 100 jackknife trials of Type I AGN realization. The blue dashed curve corresponds to a Gaussian fit, while the green long-dashed curve shows the AGND fit. The inset panel presents the same distributions on a semi-log scale, highlighting the deviations between the Gaussian model and the simulation results in the high-$N_{\rm QSO}$ regime. The deep orange shade of the inset panel shows the $1\sigma$ scatter of the CIC across the 100 jackknife trials (16th--84th percentiles). In contrast, the pale orange shade shows the full range of the scatter. Since the extreme overdense regions (e.g., $N_{\rm QSO}\geq5$ for Case~A) do not appear in every selection, the minimum probability often reaches zero. When plotted on a semi-log scale, this results in the shade having an artificially large vertical extent, as the lower bound effectively extends down to the axis limit.}
    \label{fig:count-in-cells}
\end{figure*}

\par The BHMF from our simulation (Figure~\ref{fig:BH_mass_function}) agrees well with other simulations for $\log_{10} (M_{\rm BH}/M_\odot) \gtrsim 7.0$, further confirming the robustness of our BH modeling.

\par Overall, these comparisons demonstrate that the BH population in our simulation is consistent with both observations and other state-of-the-art simulations. This provides a solid and reliable foundation for the subsequent quasar clustering analyses presented in this work.

\subsection{Count-in-cells (CIC) analysis} \label{sec:CIC}

\par  We perform a count-in-cells (CIC) analysis, a statistical method widely used to quantify the spatial distribution of objects such as galaxies \citep[e.g.,][]{ueda_counts--cells_1996,yang_galaxy_2011,shin_new_2017}. 
The CIC is a one-point statistic that characterizes the probability distribution of object counts within fixed spatial volumes $V$ (``cells''). 
In practice, the survey or simulation volume is partitioned into cells of a chosen size, typically centered on grid nodes and allowing overlap, and the number of objects $N$ within each cell is recorded. 
From the ensemble of counts, the probability distribution function $P_{\rm CIC}(N;V)$ is then constructed. 

\par The shape of $P_{\rm CIC}(N;V)$ provides a sensitive probe of cosmic structure. Deviations from a purely random (Poisson) distribution signal the presence of clustering, while the tails of the distribution capture rare overdense or underdense regions. As such, CIC analysis is a powerful tool for quantifying the clustering of objects and for detecting non-Gaussian signatures in the underlying density field, which encode key information about structure formation and the nature of gravity. 

\par We employ the CIC analysis to characterize both the shape of $P_{\rm CIC}(N;V)$ and the probability of extreme overdense regions within our simulation. In this work, we define each cell as a sub-box of size $L_{\rm cell} = 30\,h^{-1}\,{\rm Mpc} = 44.3\,{\rm Mpc}$ (for $h=0.67742$; comoving), which is close to what \citet{liang_cosmic_2025} took.  From the full simulation volume ($L=200\,h^{-1}\,{\rm cMpc}$), we randomly select $10^6$ such cells. This random sampling approach enables us to identify quasar overdensities that a purely grid-based cell placement might miss. 

\par Figure~\ref{fig:count-in-cells} presents the resulting CIC distributions, together with fitted models. We consider both a Gaussian function and the asymmetric generalized normal distribution (AGND) as fitting forms. 
We performed a fitting of the median values of the CIC distribution using the non-linear least squares method via the Levenberg-Marquardt algorithm (as implemented in the \texttt{scipy.optimize.curve\_fit} package, \citealt{2020SciPy-NMeth}). 
The best-fit parameters for the Gaussian and AGND models are summarized in Table~\ref{tab:CIC_fit_parameters}. The quoted uncertainties represent the 1$\sigma$ asymptotic standard errors derived from the diagonal elements of the covariance matrix.

\begin{table*}[htpb]
    \centering
    \caption{Best-fit parameters of the Gaussian model and the asymmetric generalized normal distribution (AGND) model for the quasar CIC PDF.}
    \label{tab:CIC_fit_parameters}
     \begin{tabular}{ccc} \hline \hline
        Gaussian & $\mu$ & $\sigma$ \\ \hline
        Case~A & $0.14 \pm 0.042$ & $0.48 \pm 0.020$\\
        Case~B & $0.94 \pm 0.075$  & $1.36 \pm 0.058$ \\ \hline
    \end{tabular}
    \begin{tabular}{cccc} \hline \hline
        AGND & $\alpha$ & $\kappa$ & $\xi$ \\ \hline
        Case~A & $0.525 \pm 0.0014$ & $0.54 \pm 0.016$ & $0.041 \pm 0.0027$ \\
        Case~B & $1.37 \pm 0.020$ & $0.44 \pm 0.041$  & $1.217 \pm 0.032$ \\ \hline
    \end{tabular}
    
\end{table*}

\par The Gaussian model is characterized by two parameters: mean and standard deviation.
In contrast, the AGND model naturally accommodates skewness and asymmetry in the distribution, providing a better representation of the CIC results, as demonstrated in Figure~\ref{fig:count-in-cells}.
The AGND model is specifically designed to capture non-Gaussian features in probability distributions and has previously been employed for CIC analyses by \citet{shin_new_2017}. We adopt the notation $\mathcal{N}_{\rm v2}$ following their designation of this function as the `generalized normal distribution version 2'. Its functional form is
\begin{align}
  \mathcal{N}_{\rm v2} (x) = \frac{\mathcal{N}(y)}{\alpha + \kappa (x - \xi)}\, ,
\end{align}
\begin{equation}
    y = 
    \begin{cases}
     \frac{1}{\kappa} \ln \qty[1 + \frac{\kappa (x - \xi)}{\alpha}] & (\kappa \neq 0) \, ,\\
     \frac{x - \xi}{\alpha} & (\kappa = 0) \, ,
\end{cases}
\end{equation}
where $\mathcal{N}(y)$ denotes the Gaussian distribution. 
The three free parameters have intuitive interpretations: 
$\alpha$ sets the overall width of the distribution, 
$\xi$ specifies its location (analogous to the median), 
and $\kappa$ controls the degree of asymmetry. 
This flexibility allows the AGND to reproduce the skewed, heavy-tailed shapes that characterize the CIC distributions in our simulation and to provide a physically meaningful alternative to Gaussian fits. In passing, we note that the generalized extreme value (GEV) distribution has also been employed in cosmological one-point statistics \citep[e.g.][]{thompson_rise_2015, repp_precision_2018}, where it is motivated by extreme-value theory; however, in contrast to such approaches that focus on block maxima, our analysis models the full parent CIC distribution.

\par As shown in Figure~\ref{fig:count-in-cells}, the Gaussian fit fails to reproduce both the skewness and the high-density tail of the CIC distribution in both Cases. This visual inadequacy is quantitatively confirmed by a likelihood ratio test. To account for the cell-to-cell correlations in our random sampling, we define the log-likelihood by scaling with the effective number of independent cells, $N_{\rm eff} = V_{\rm sim}/V_{\rm cell} = (200 \, h^{-1} \, {\rm Mpc}/30 \, h^{-1} \, {\rm Mpc})^3 \simeq 296$:
\begin{align}
    \ln \mathcal{L} = N_{\rm eff} \sum_i P_{\rm CIC, sim} (N_i; L_{\rm cell}) \ln P_{\rm CIC, model} (N_i; L_{\rm cell}) \, .
    \label{eq:log_likelihood_scaled}
\end{align}
Using this formulation, we perform a likelihood ratio test to evaluate the null hypothesis that the Gaussian model provides a sufficient description of the distribution. We calculate the test statistic $D = -2 (\ln \mathcal{L}_{\rm Gauss} - \ln \mathcal{L}_{\rm AGND})$, which follows a $\chi^2$ distribution with one degree of freedom, corresponding to the difference in the number of free parameters between the two models. The resulting $p$-values are $\sim 10^{-29}$ for Case~A and $\sim 10^{-31}$ for Case~B, allowing us to decisively reject the Gaussian null hypothesis in favor of the AGND model.
Furthermore, based on the Gaussian fit results, regions corresponding to $12\sigma_{\rm Gauss}$ overdensities (representing the highest overdensities observed in both Case~A and Case~B) are found to appear naturally in the simulation. While both samples reach this same statistical significance level relative to the Gaussian fit, the corresponding physical overdensities differ markedly: Case~A reveals extremely dense regions reaching $\delta_{\rm QSO} \sim 23$, whereas Case~B exhibits a comparatively moderate maximum of $\delta_{\rm QSO} \sim 10$. This discrepancy suggests the possibility that sample selection effects may enhance the visibility of such extremely high-density regions. These findings highlight the inadequacy of the Gaussian assumption in the extreme tail. In contrast, the AGND model provides an excellent fit across the full range of CIC counts. This result demonstrates that Gaussian fitting is inadequate for evaluating quasar overdensities in the high-density regime, whereas the AGND model offers a more accurate and physically motivated description of the underlying distribution. Because \citet{liang_cosmic_2025} evaluated the CH overdensity using a Gaussian fit, this comparison is essential for reassessing the claimed statistical significance of the CH. This result also underpins our later conclusion (Section~\ref{sec:discussion}) that the CH is more probable within $\Lambda$CDM than implied by a Gaussian-based analysis.

\subsection{NND analysis of quasars} \label{sec:NND}

\begin{figure*}[htpb]
    \centering
    \begin{minipage}{.5\linewidth}
        \includegraphics[width=\linewidth]{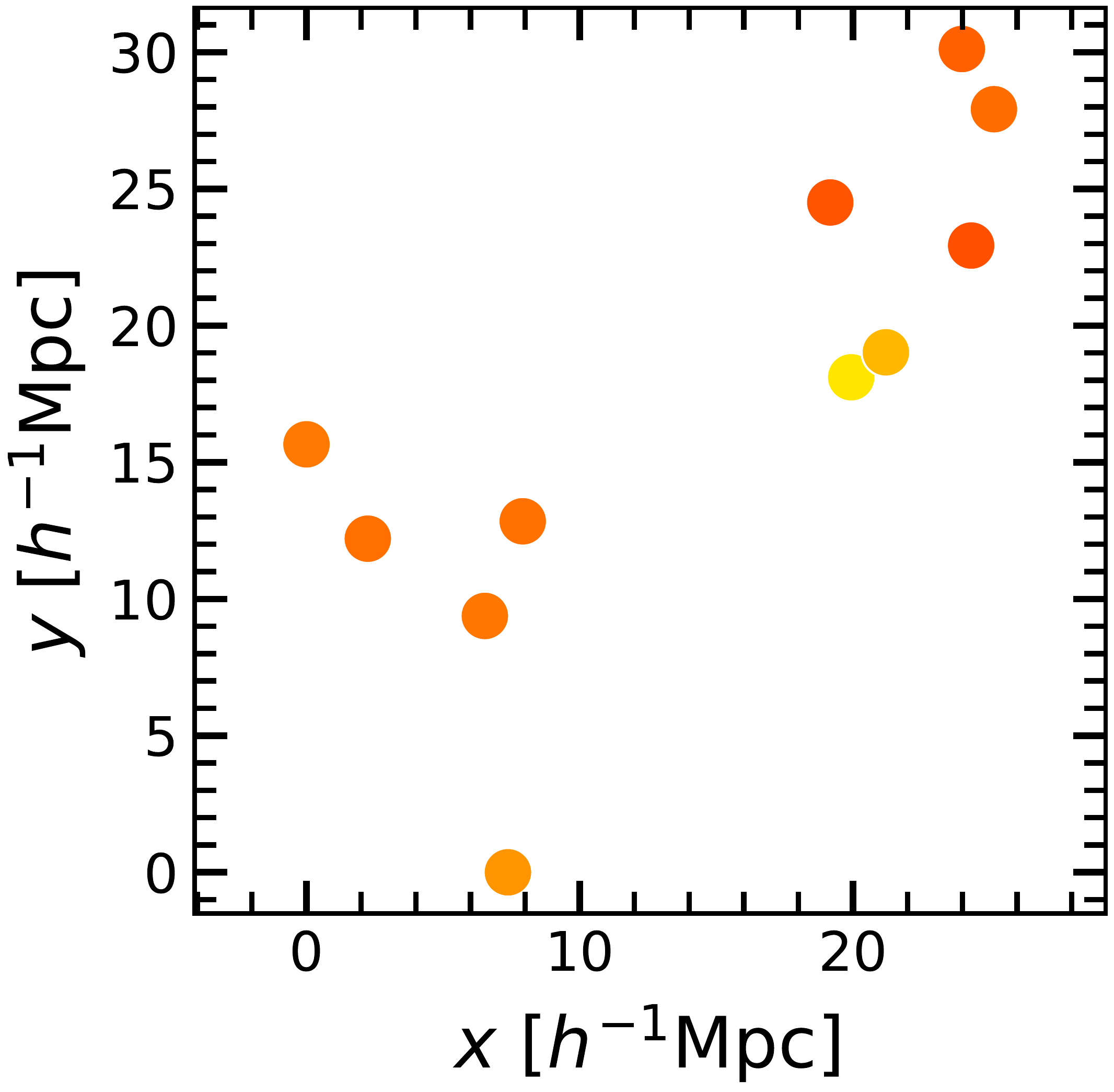}
        \subcaption{The Cosmic Himalayas}
        \label{fig:quasar_distribution_CH}
    \end{minipage}
    \\
    \begin{minipage}[t]{0.44\linewidth}
        \includegraphics[width=\linewidth]{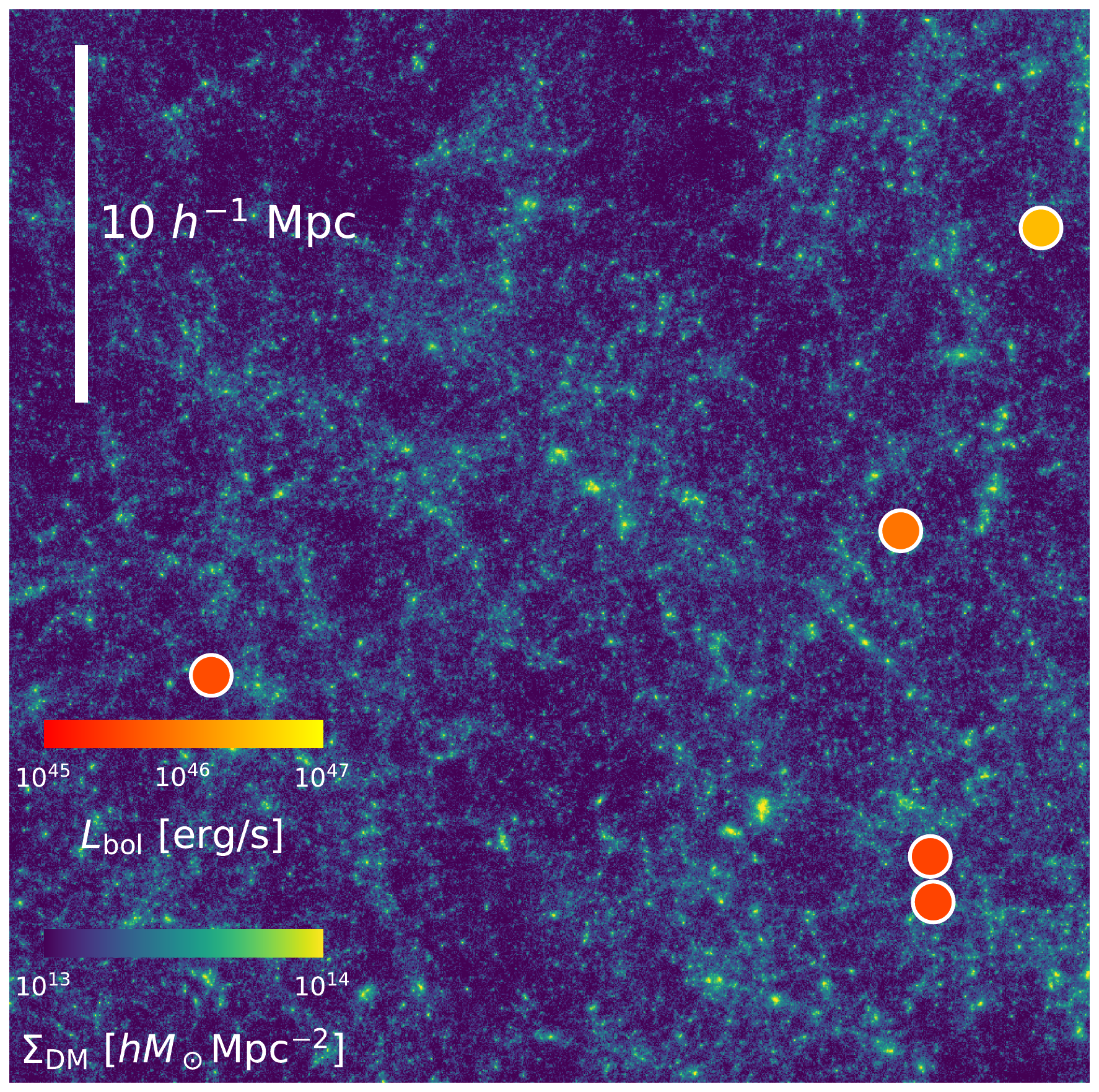}
        \subcaption{Case~A}
        \label{fig:quasar_distribution_CaseA}
    \end{minipage}
    \begin{minipage}[t]{.44\linewidth}
        \includegraphics[width=\linewidth]{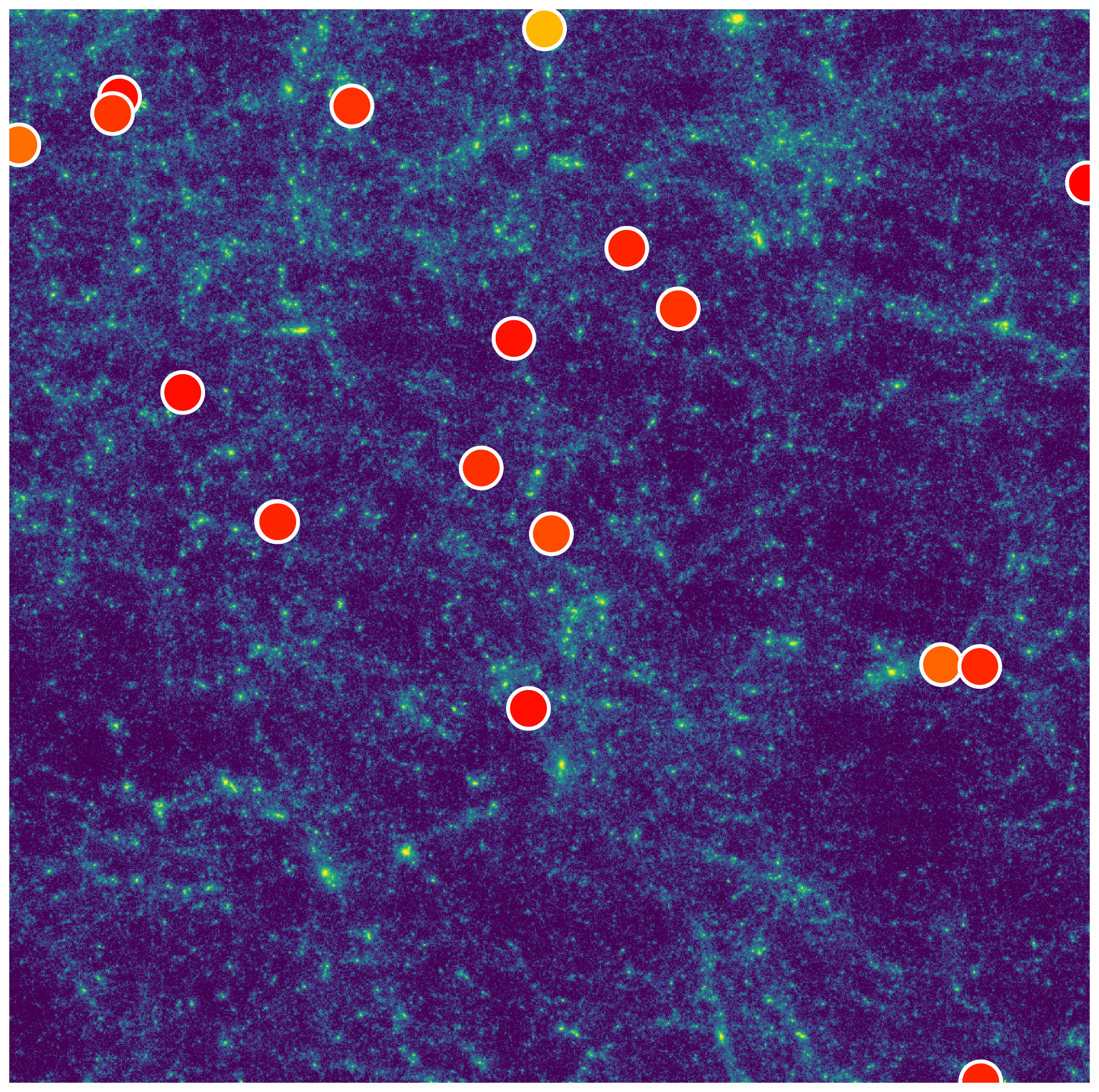}
        \subcaption{Case~B}
        \label{fig:quasar_distribution_CaseB}
    \end{minipage}

    \caption{Projection maps of quasar distributions. \\ 
    {\it Panel (a)}: The observed quasar distribution in the Cosmic Himalayas (CH) \citep{liang_cosmic_2025}, shown in the comoving frame where $(x, y)$ represents the comoving coordinates relative to the fiducial point. This $\sim (30 \, h^{-1} \, {\rm Mpc})^3$ region contains 11 quasars. Points represent quasars, color-coded by bolometric luminosity, using the same normalization and color scale as in panel (b). \\
    {\it Panel (b)}: An example of a candidate region for a CH analog in the Case~A selection (containing 5 quasars). This distribution corresponds to the most overdense cell identified in a single random realization of the Type~I AGN selection. Points represent quasars, color-coded by their bolometric luminosity calculated from Equation~\eqref{eq:luminosity-accretionrate}, and the background color shows the projected dark matter column density. \\
    {\it Panel (c)}: An example of a CH analog in the Case~B selection (containing 17 quasars), similarly extracted from a single realization of the Type~I selection. The color coding and background are identical to those in panel (b).}
    \label{fig:quasar_distribution}
\end{figure*}

\begin{figure*}       
    \begin{minipage}[t]{0.395\linewidth}
         \includegraphics[width=\linewidth]{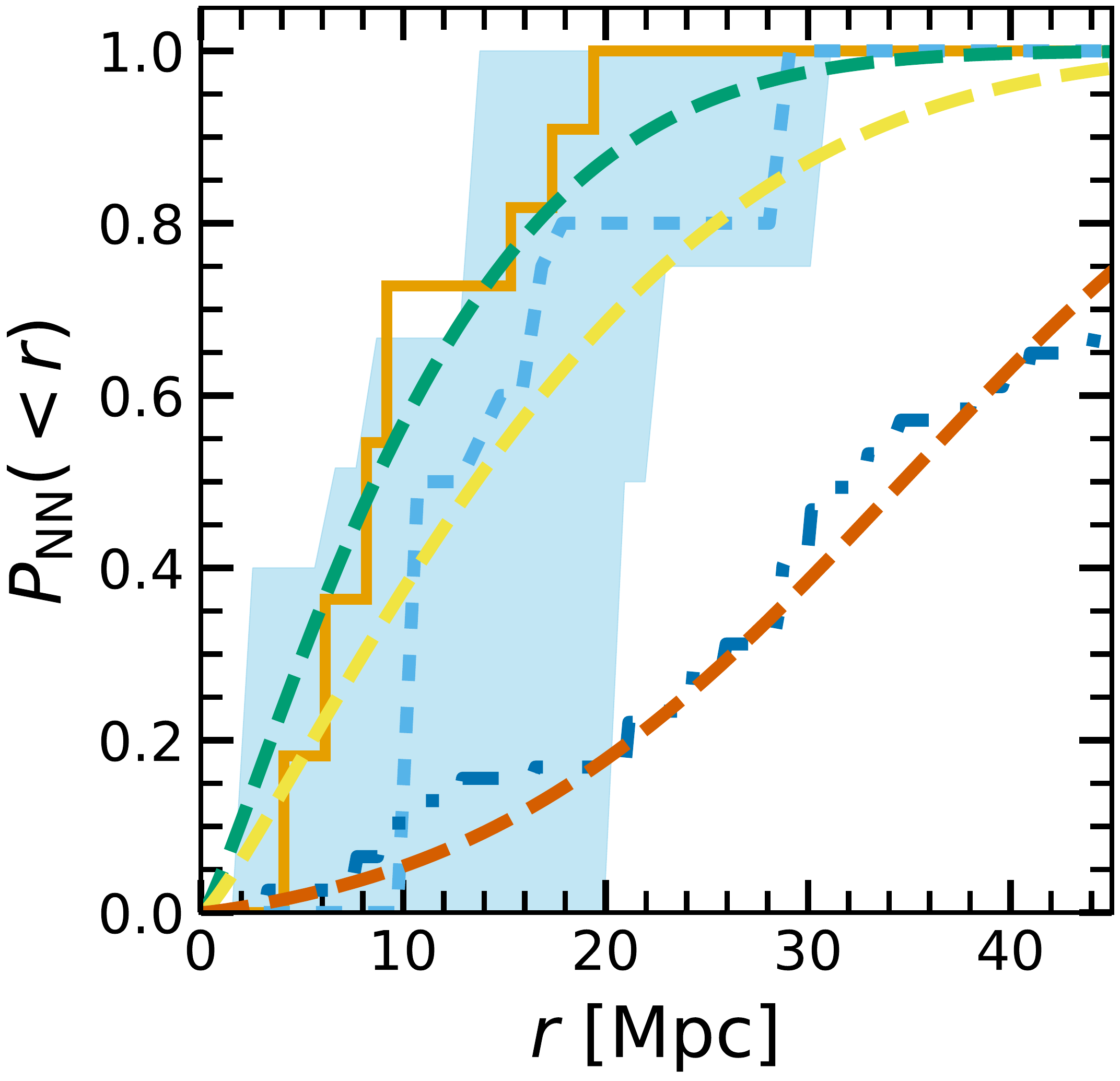}
         \subcaption{NND analysis for Case~A.}
    \end{minipage}
    \begin{minipage}[t]{0.6\linewidth}
        \includegraphics[width=\linewidth]{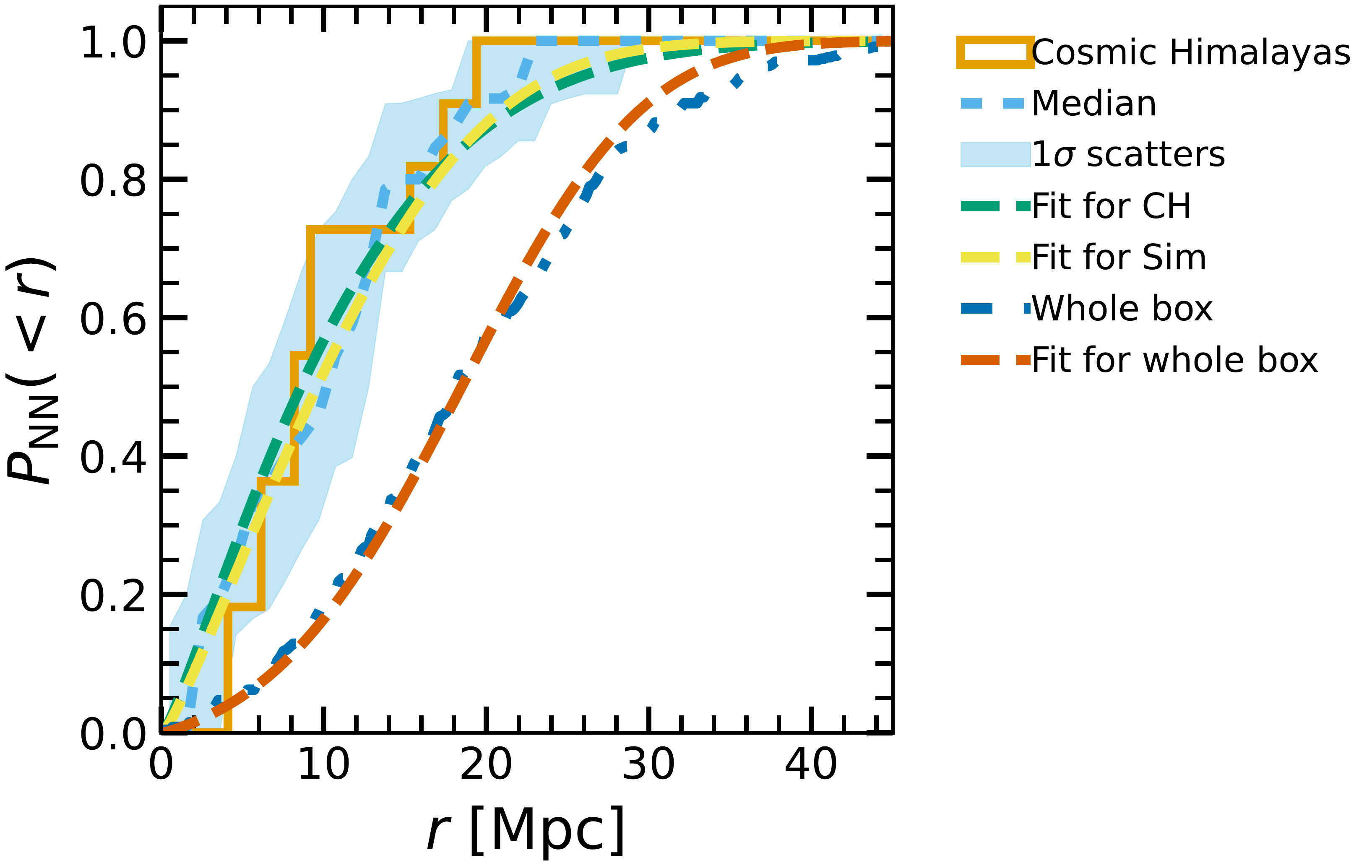}
        \subcaption{NND analysis for Case~B.}
    \end{minipage}
    \caption{Nearest-neighbor distribution (NND) analysis results. The orange line shows the NND of quasars in the CH. The sky-blue line with shaded region represents the NND of the most overdense quasar regions, identified through CIC analysis in each Type-I AGN sample realization for each selection case (\texttt{BF50NEL}). Specifically, the associated shade indicates the $1\sigma$ scatter obtained by the 100 jackknife trials(16th--84th percentile). 
    The blue dotted line shows an example of the NND calculated for the entire $200 \, h^{-1} \, {\rm Mpc}$ simulation box in a trial.
    At first glance, the NND of the CH appears to show excellent agreement with Case~B, whereas Case~A shows a clear deviation. 
    However, these apparent similarities and deviations stem primarily from differences in the mean quasar number density rather than intrinsic clustering signals. To address this, we quantify the intrinsic clustering strength (effective correlation length $r^{\rm eff}_0$) by fitting the NND using the formulation derived in Equation~\eqref{eq:relation_NND_TPCF}. 
    The green, yellow, and red dotted lines present the fitting results of the NND via a single power-law model of the TPCF $\xi(r) = (r/r^{\rm eff}_0)^{-\gamma}$, with a fixed slope of  $\gamma = 1.8$. Specifically, the green line corresponds to the fitting result for the CH, the yellow line for the most overdense regions in our simulation, and the red line for the NND of the whole simulation box.  The correlation lengths by these fittings are summarized in Table~\ref{tab:correlation_length}.
    } 
    \label{fig:NND_quasar}
\end{figure*}

\begin{table*}[t]
    \centering
    \caption{Effective correlation length $(r_0^{\rm eff})$ estimated by the fitting of NND.}
    \begin{tabular}{cccc} \hline \hline
        $r^{\rm eff}_0$ [$h^{-1}\,$Mpc] & CH & Case~A & Case~B \\ \hline
         Extreme overdensities & $ 29.4 \pm 0.42$  & $30.9 \pm 0.56$ & $13.69 \pm 0.071$ \\ \hline
         whole box & -- & $8.4 \pm 0.35$ & $4.97 \pm 0.068$ \\ \hline
    \end{tabular}
    \label{tab:correlation_length}
\end{table*}

\par As a next step for quantifying quasar clustering, we employ the nearest-neighbor distribution (NND). The NND is defined as the probability distribution of the distance $r$ from an object to its closest neighbor. It can be characterized by the cumulative distribution function (CDF), $P_{\rm NN}(<r)$, which gives the probability that the nearest neighbor lies within distance $r$. The corresponding probability density function (PDF) is $p_{\rm NN}(r) = \dfrac{\dd P_{\rm NN}(<r)}{\dd r}$.
An excess probability at small separations---i.e., a higher $P_{\rm NN}(<r)$ at small $r$ relative to a random distribution---indicates a population that is more strongly clustered. 

\par To quantify clustering in localized regions, we focus on the NND calculated for the most quasar-overdense cell identified by the CIC analysis for each Type I AGN sample realization. 
Figure~\ref{fig:quasar_distribution} presents the projection maps of the quasar distributions. Panel (a) shows the distribution in the CH, while panels (b) and (c) display examples of a single Type I AGN sample realization for Case~A and Case~B, respectively. The NND was then calculated for the quasars within these extreme overdensities. However, a direct comparison of NNDs across samples and the observation (CH) is problematic because NND fundamentally depends on the number density of the target objects, which varies significantly across datasets. 


\par To overcome this limitation and enable a robust comparison, we model $P_{\rm NN}(<r)$ by incorporating a power-law two-point correlation function (TPCF; $\xi(r)$),  
which describes the excess probability of finding a pair at a separation $r$. 
The probability of {\it NOT} finding a pair within a distance $r$ denoted as $P_{\rm NN}(>r)$, is related to the TPCF by the following expression \citep{Chiu_Stochastic_2013}: 
\begin{align}
    P_{\rm NN}(>r) = \exp(- \int_0^r 4\pi r^2 \bar{n} (1+\xi(r)) \dd r)\, ,
\end{align}
where $\bar{n}$ is the cosmic mean number density of objects. Consequently, the cumulative NND $P_{\rm NN}(<r)$ is given by: 
\begin{align} 
    P_{\rm NN}(<r) &= 1 - P(>r)  \\
    \label{eq:relation_NND_TPCF}
    &= 1 - \exp(- \int_0^r 4\pi r^2 \bar{n} (1+\xi(r)) \dd r)\,.
\end{align}
Using this relation, we can isolate the clustering signature (encoded in $\xi(r)$) from the number density effects. 

\par To quantify the clustering strength using this formulation, we fit this analytical model to the observed $P_{\rm NN}(<r)$ from our simulations and the observation (CH), using a single power-law TPCF, $\xi(r) = (r/r_0)^{-\gamma}$, with the power-law index fixed to the standard value of $\gamma = 1.8$. This approach allows us to determine the correlation length $r_0$, which effectively quantifies the clustering strength independent of the sample density. 

\par The aforementioned relations between $P_{\rm NN}$ and $\xi(r)$ strictly hold only for Gaussian-Poisson processes, as $P_{\rm NN}$ inherently contains information from higher-order $N$-point correlation functions. In non-Gaussian point processes, however, the parameter $r_0$ can be replaced by an effective parameter $r_0^{\rm eff}$. This effective parameter accounts for contributions from higher-order correlations and serves as a robust indicator of intrinsic clustering. 

Figure~\ref{fig:NND_quasar} shows the NND analysis results for Case~A and Case~B. At first glance, the NND of the CH appears to show excellent agreement with Case~B, whereas Case~A shows a clear deviation. However, we must caution that a direct comparison of $P_{\rm NN}(<r)$ is misleading because the cosmic mean number density of quasars varies significantly between samples. 
To address this, we estimate the effective correlation length $r_0^{\rm eff}$, by fitting the observed $P_{\rm NN}(<r)$ with the theoretical formulation given in Equation~\eqref{eq:relation_NND_TPCF}. The fitting is performed using the non-linear least-squares algorithm implemented in \texttt{scipy.optimize.curve\_fit} \citep{2020SciPy-NMeth}, where $r_0$ in Equation~\eqref{eq:relation_NND_TPCF} is treated as a free parameter. This approach allows us to decouple the effect of the mean number density and quantify the intrinsic clustering strength of each sample.

\par Through this fitting procedure, we obtained the correlation lengths and their $1\sigma$ errors listed in Table~\ref{tab:correlation_length}. Notably, our analysis reveals that the apparent agreement between Case~B and the CH is driven primarily by the higher number density in the Case~B selection rather than by intrinsic clustering properties. In contrast, when correcting for density effects, the correlation length derived for the Case~A selection $(r^{\rm eff}_0\simeq 30 \, h^{-1} \, {\rm Mpc})$ is consistent with that of the CH. Regarding the similarities between the simulated cases, both Case~A and Case~B exhibit significant spatial variations in clustering strength; depending on the specific cell selection, the derived $r^{\rm eff}_0$ can vary by a factor of $\sim 3$ relative to the global trend. These findings support our CIC results, suggesting that the strong clustering signals similar to the CH naturally emerge from the combined effects of sample selection biases and cosmic variance.

\section{Discussion} \label{sec:discussion} 
To clarify the origin of the extreme peak-height value ($\nu = \delta/\sigma = 16.9$) argued for CH, we performed the CIC analyses on our simulation samples. 
It is important to highlight the methodological differences between our analysis and that of \citet{liang_cosmic_2025}. First, our simulation box has a size of $L_{\rm box}=200\,h^{-1}\,{\rm Mpc}\,\simeq\,295\,{\rm Mpc}$, whereas the SDSS observational fields span Gpc-scale volumes. Second, \citet{liang_cosmic_2025} divided the survey into a fixed grid, which risks splitting intrinsically dense regions across multiple cells and diluting the signal. By contrast, we considered both $(L_{\rm box}/L_{\rm cell})^3 \approx (200/30)^3 \sim 3\times10^2$ independent non-overlapping cells and overlapping random placements to obtain a smoother estimate of the PDF shape. 
This approach is well-suited to constraining the functional form of the CIC distribution, though it cannot by itself capture the survey-wide tail statistics expected in Gpc-scale fields.

The fundamental mode of the box is $k_f = 2\pi / L_{\rm box} \approx 0.031\,h\,\mathrm{Mpc}^{-1}$, comparable to $1/L_{\rm cell}$. Thus, the modes that dominate fluctuations within a $30\,h^{-1}\,\mathrm{Mpc}$ cell---which set the skewness and tail of the one-point distribution---are well sampled by many higher-$k$ modes inside the box. In other words, the $200\,h^{-1}\,\mathrm{Mpc}$ volume is sufficient to determine the \emph{shape} of $P_{\rm CIC}(N_{\rm QSO};L_{\rm cell})$, even if it cannot measure the absolute rarity of the most extreme cells across a Gpc-scale survey.

\par On Gpc-scale volumes, even mild non-Gaussianity in the true CIC probability distribution function can make a ``$\sigma$-language'' summary a misleading exaggeration of the PDF tail. Our simulation shows that the quasar CIC is distinctly non-Gaussian;  a Gaussian with a fixed mean fails to reproduce the high-$N_{\rm QSO}$ tail, whereas the AGND model provides an excellent fit, consistent with the results for dark matter haloes reported by \citet{shin_new_2017}. Thus, the reported ``$16.9\sigma$'' significance corresponds to an incorrect likelihood model, and $\sigma$-equivalents lose their interpretive meaning when the assumed likelihood is wrong. We explicitly demonstrated this Gaussian mis-fit and the successful AGND fit in Figure~\ref{fig:count-in-cells}.

\par A more appropriate metric is the exceedance probability under the fitted non-Gaussian PDF:
\begin{equation}
    p_{\geq N_*} \;=\; \sum_{N=N_*}^{\infty} P_{\mathcal{N}_{\rm v2}}(N; L_{\rm cell}) \, ,
\label{eq:pcell}
\end{equation}
where $N_*$ serves as a threshold count. This metric $p_{\geq N_*}$ represents the probability of finding $N_{\rm QSO} \geq N_*$ quasars within a single cell. This probability can then be converted into a survey-level probability using the effective number of independent cells, $\zeta$:
\begin{equation}
    P_{\rm exist}(N_{\rm QSO}{\geq N_*}, V_{\rm survey}) \;=\; 1 - \bigl(1 - p_{\geq N_*}\bigr)^{\zeta}, 
\label{eq:psurvey}
\end{equation}
where $\zeta \approx (V_{\rm survey} / L_{\rm cell}^3 )$.

In other words, 
$p_{\geq N_*}$ given in Equation~\eqref{eq:pcell}, is the cell-level exceedance probability that one random cell of size $L_{\rm cell}$ has at least $N_*$ quasars; i.e., it tells us how rare such a cell is locally. 
The latter term in Equation~\eqref{eq:psurvey}, $\bigl(1 - p_{\geq N_*}\bigr)^{\zeta}$, represents the probability that none of the $\zeta$ cells exceeds the threshold. 
Then, Equation~\eqref{eq:psurvey} represents the complement, i.e., the probability that at least one cell in the survey exceeds the threshold $N_*$.  This indicates the likelihood that the survey will contain at least one such cell, given its vast volume.


We first applied this formulation to the Case A samples, which exhibit a mean quasar number density similar to the SDSS survey \citep{liang_cosmic_2025}. We set the threshold to $N_* = 8$, which corresponds to the highly overdense regime of $\delta_{\rm QSO} \gtrsim 17\sigma_{\rm Gauss}$. With the parameters set as $L_{\rm cell}=30 \ h^{-1} {\rm Mpc}$, and a survey volume of $V_{\rm survey} = 1 \, h^{-3}\,{\rm Gpc}^3$, the probability of finding at least one such cell is $P_{\rm exist} = 0.71$, computed using the fitted AGND parameters in Table~\ref{tab:CIC_fit_parameters}, with $p_{\geq N_*}\approx 3.306\times 10^{-5}$ and $\zeta=37,037$.
When the survey volume is increased to $V_{\rm survey} = (2 \, h^{-1} \, {\rm Gpc})^3$, it rises to $P_{\rm exist} \approx 1$, with $\zeta  = 296,296$ and the same value of $p_{\geq N_*}$.
\par Next, we examined the Case~B samples. We set the threshold to $N_* =25$, which again corresponds to the extreme overdensity level of $\delta \gtrsim 17 \sigma_{\rm Gauss}$. With the same cell size ($L_{\rm cell}=30 \ h^{-1} {\rm Mpc}$) and a survey volume of $V_{\rm survey} = 1 \, h^{-3}\,{\rm Gpc}^3$, it yields a probability of $P_{\rm exist}  = 0.024$ with $p_{\geq N_*}\approx 6.5316\times 10^{-6}$. Increasing the survey volume to $V_{\rm survey} = (2 \, h^{-1} \, {\rm Gpc})^3$ increases this probability to $P_{\rm exist}  = 0.18$. 

While a more detailed treatment of the quasar selection function is beyond the scope of this paper, these results strongly support the notion that the emergence of extreme overdense regions, such as the CH, is a natural consequence in a Gpc-scale survey like SDSS. Crucially, the clear difference in the realization probability between Case~A and Case~B demonstrates that the selection criteria significantly influence the observed emergence of these high-density regions.


\par Finally, we acknowledge that this study has focused primarily on the statistical properties of quasar distributions. We have not explicitly investigated the underlying dark matter environments of the identified CH analogs to determine whether they physically reside within genuine protoclusters. Consequently, while we confirm that the extreme quasar overdensity is statistically consistent with $\Lambda$CDM, it remains an open question whether the CH explicitly traces the most massive dark matter structures at this redshift. A detailed examination of the halo environments hosting these CH analogs is reserved for future work. We stress that our conclusions rely on statistical properties common to biased tracers in $\Lambda$CDM and are validated using both hydrodynamic (\textit{CROCODILE}) and dark-matter-only (\textit{Uchuu}, Appendix \ref{appendix:Uchuu}) simulations with independent tracer constructions.

\par In summary, this study provides a crucial revision to the conventional assessment of extreme overdensities. We have demonstrated that the Gaussian model is fundamentally inadequate for quantifying quasar overdensities in the high-density regime because it fails to capture the significant skewness and heavy-tailed nature of the underlying CIC distribution. To resolve this, we employed a novel framework based on the AGND model. This framework is robust because it (i) respects the correct heavy-tailed form measured directly in our simulations, and (ii) scales naturally to the Gpc-scale volume of the SDSS survey. 
Crucially, our NND analysis further supports this conclusion by revealing that regions exhibiting two-point clustering properties highly analogous to the CH (termed CH analogs) naturally emerge within our simulation volume. This finding suggests that the observed strong clustering signal is not an anomalous effect but rather a plausible consequence of the combined effects of sample selection biases (as modeled by our Case A and Case B criteria) and the inherent cosmic variance. Applying this robust method reveals that the rarity of the observed extreme structure, such as the CH, is substantially lower than the often-cited Gaussian-based `$\sim 17\sigma$' claim. By accurately modeling the underlying distribution, our results successfully reconcile the existence of the CH with expectations from the standard $\Lambda$CDM paradigm.

\section{Summary} \label{sec:summary}
\par In this paper, we investigated the emergence and properties of extreme quasar overdensities using the {\it CROCODILE} cosmological hydrodynamic simulations, examining them from both large-scale (CIC) and small-scale (NND) perspectives. Our simulation provides BH masses and accretion rates, from which we estimated quasar luminosities. The resulting BH population is consistent with those of other simulations and observations of Eddington ratios, X-ray luminosity, and BH mass function (Figures~\ref{fig:lambda_Edd_distribution}--\ref{fig:BH_mass_function}). Based on these samples, we selected quasars according to mass- and luminosity-based criteria (Cases~A and B) and analysed their clustering properties. 

From our CIC analysis, we find that the observed distribution of quasar counts deviates strongly from the Gaussian assumption, particularly in the high-density tail. As a consequence, conventional Gaussian fits systematically overestimate the rarity of extreme overdensities (e.g., inferring $\delta \sim 12 \sigma_{\rm Gauss}$ for simulated overdense cells that are actually more common), whereas a non-Gaussian functional form, such as the AGND model, provides a significantly more accurate description across the full range of counts. Both Case~A and B highlight this effect, with the more biased selection (Case~A) exhibiting the largest discrepancy. We reinforced this finding by calculating the exceedance probability using the AGND model in both the {\it CROCODILE} simulation (Sec~\ref{sec:discussion}) and the {\it Uchuu} simulation (appendix~\ref{appendix:Uchuu} which has a much larger boxsize of $2 \, h^{-1} \, {\rm Gpc}$. Our CIC results therefore imply that the CH overdensity has been substantially overstated. 


\par For further investigation into the physical origin of the strong clustering, we perform the NND analysis, which allows us to quantify the two-point statistics highly localized, small regions. By connecting the NND with TPCF, we find that the strong clustering signature observed in the CH ($r^{\rm eff}_0 \sim 30 \, h^{-1} \, {\rm Mpc}$) naturally arises in samples from Case~A, which represents a more biased sample selection. This result strongly supports the statement that the extreme clustered regions of quasars emerge from the compound effect of sample selection biases and the spatial dependence of the observed field, thus implicating sample selections and cosmic variance as the primary drivers. 

\par In summary, the apparent tension between the existence of extreme structures (like CH analogs) in $\Lambda$CDM simulations and the extreme overdensity reported by \citet{liang_cosmic_2025} stems from the inappropriate use of a Gaussian assumption for the CIC distribution. Once the correct non-Gaussian functional form is adopted, the inferred rarity of the Cosmic Himalayas is substantially reduced, allowing the simulation adopting the standard universe model to successfully observe these extreme overdensities. Complementing this, our NND analysis confirms that strong clustering signals can be observed due to the compound effect of sample selection biases (as modeled by Case~A and Case~B) and cosmic variance. These findings conclusively support our conclusion that the extreme overdensity and clustering of quasars can be naturally produced within the $\Lambda$CDM framework.

\begin{acknowledgments}
This work used computational resources of the SQUID provided by the D3 Center of the University of Osaka, through the HPCI System Research Project (Project IDs: hp240141, hp250119). This work is partially supported by the MEXT/JSPS KAKENHI Grant Numbers 22K21349, 24H00002, 24H00241, 25K01032 (K.N.) and 21H04489 (H.Y.) and JST FOREST Program Grant Number JP-MJFR202Z (H.Y.).

\end{acknowledgments}

\FloatBarrier

\appendix

\section{Count-in-cells (CIC) analysis in the {\it Uchuu} simulation}
\label{appendix:Uchuu}

To validate the robustness of our findings based on the {\it CROCODILE} simulation box, we carried out a complementary CIC analysis using the cosmological $N$-body simulation {\it Uchuu} \citep{ishiyama_uchuu_2021}, which spans a comoving volume of $L_{\rm box} = 2\,h^{-1}\,{\rm cGpc}$. 
The {\it Uchuu} simulation provides an ideal large-volume counterpart for testing whether the functional form and tail behavior of the CIC distribution inferred from the {\it CROCODILE} box remain valid at Gpc scales, although we need a model to assign quasars to dark matter haloes as we describe below.



\par
\begin{figure}
    \centering
    \begin{minipage}[t]{0.45\linewidth}
        \includegraphics[width=\linewidth]{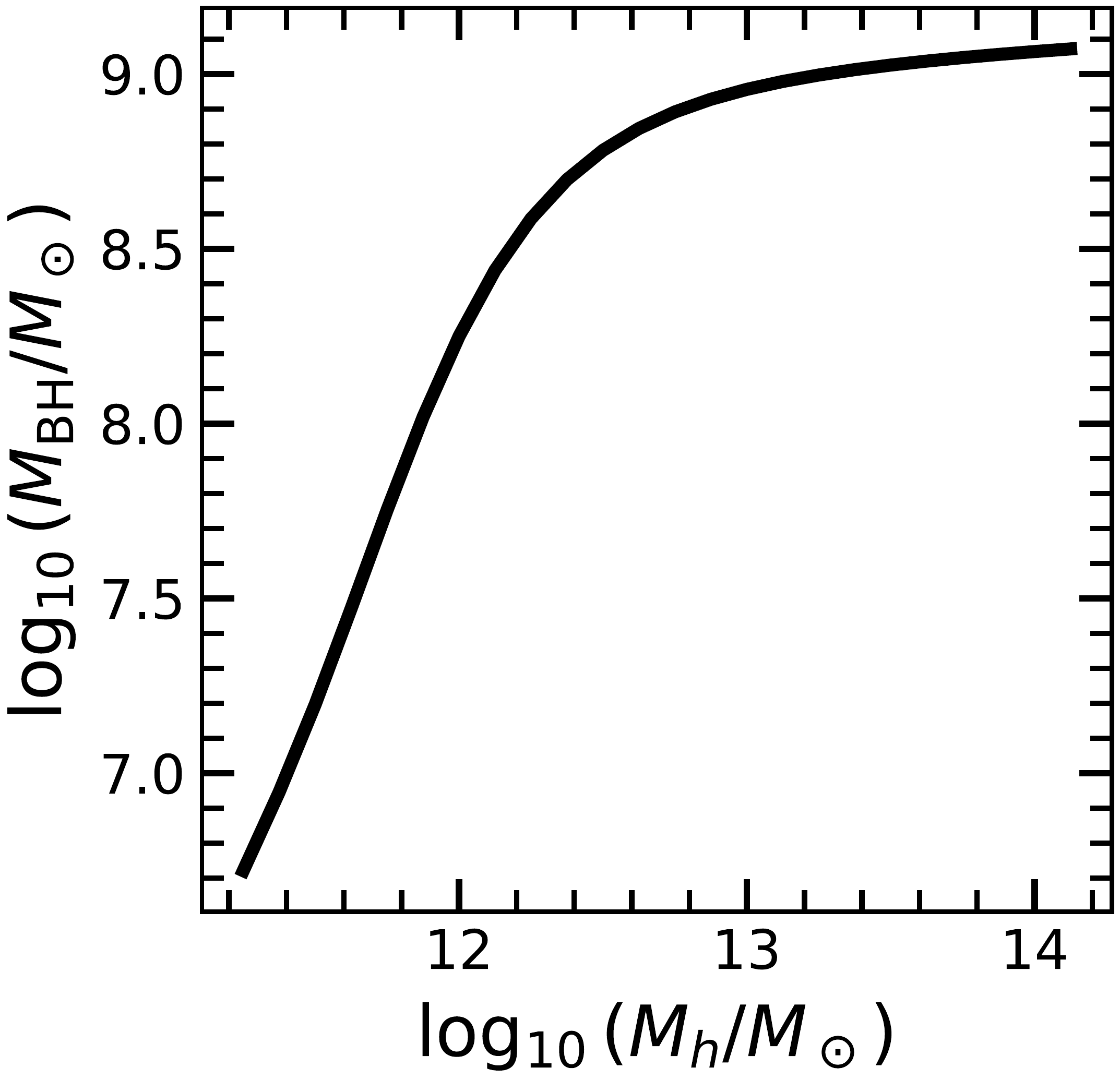}
        \caption{BH mass–halo mass relation at $z \sim 2$. The relation is derived by combining the $M_*$--$M_{\rm h}$ relation from \citet{behroozi_average_2013} at $z \sim 2$ with the $M_*$--$M_{\rm BH}$ relation from \citet{zhang_stellar_2023}.}
        \label{fig:BH_halo_mass_relation}
    \end{minipage}
    \hspace{0.05\linewidth}
    \begin{minipage}[t]{0.45\linewidth}
        \includegraphics[width=\linewidth]{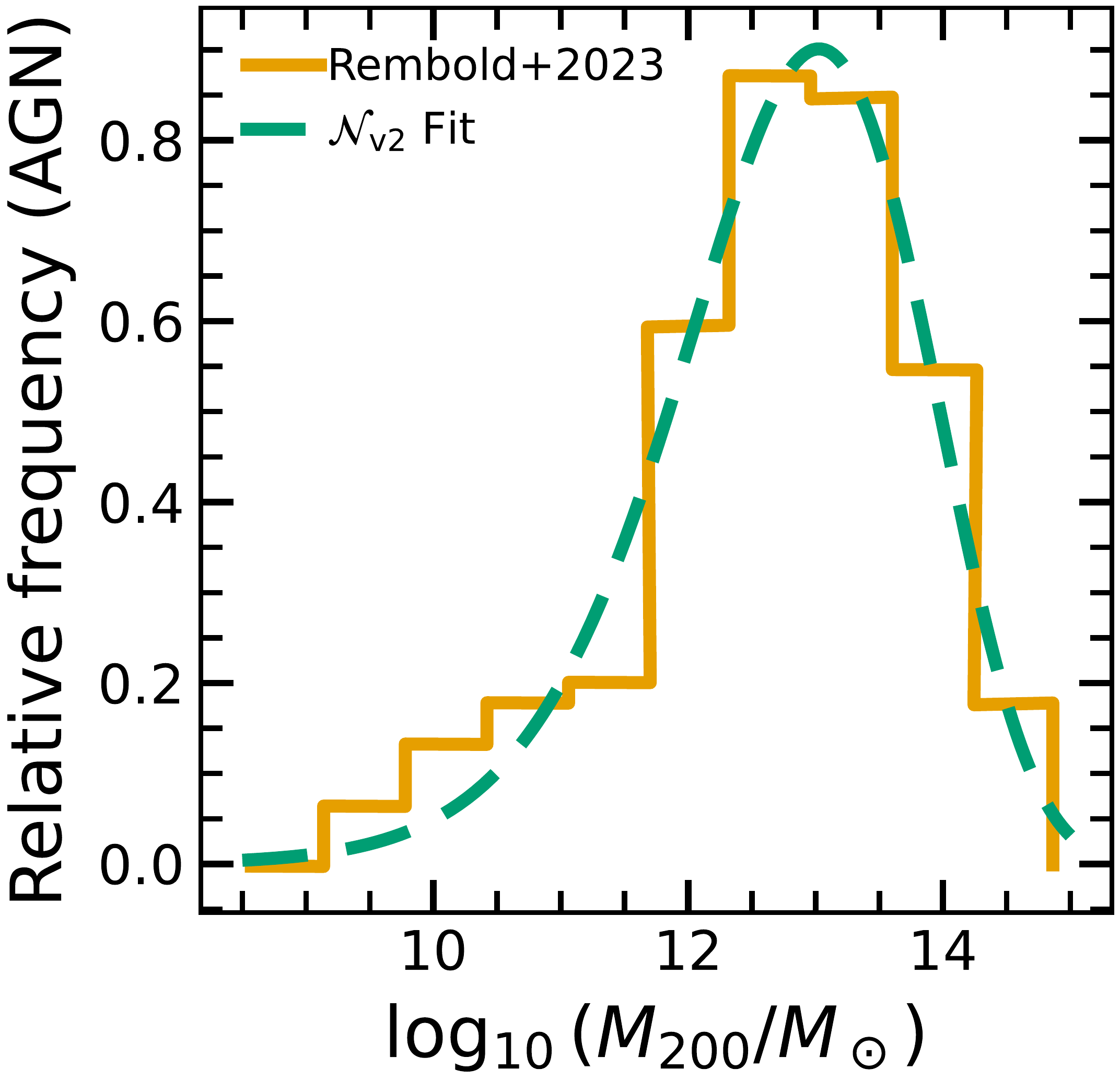}
        \caption{Relative AGN frequency as a function of halo mass. The orange histogram shows the observational results from \citet{rembold_sdss-iv_2023}, while the green dashed line represents the AGND fit to this relation. This empirical relation was used to downsample the halo population in Case~C to match the observed AGN occupation fraction.}
        \label{fig:AGN_frequency}
    \end{minipage}
\end{figure}

 In Case~C, we select halos with $\log_{10} (M_{200c}/M_\odot) > 11.5$, corresponding to typical hosts of black holes in the range $7.1 < \log_{10} (M_{\rm BH}/M_\odot) < 9.9$, consistent with our Case~B selection.  The mapping between halo mass and BH mass was derived from the $M_{\rm BH}-M_*$ relation of \citet{zhang_stellar_2023} combined with the $M_{\rm h}$--$M_{*}$ relation at $z=2.0$ from \citet{behroozi_average_2013}, as illustrated in Figure~\ref{fig:BH_halo_mass_relation}.
For this halo sample, we incorporate the observed dependence of AGN host occurrence on halo mass \citep{rembold_sdss-iv_2023}, as shown in Figure~\ref{fig:AGN_frequency}. 
The AGN fraction is modeled as a function of halo mass using an AGND model, whose amplitude was scaled such that the resulting sample yields a mean halo count per cell of $\bar{N}_{\rm halo} \simeq 0.36$. 
This fitting function is applied as a probability weight to downsample the halo catalog, ensuring that the resulting halo population reproduces the observed AGN occupation fraction. 
The best-fit AGND parameters of the observed AGN occupation fraction are $\alpha = 2.25 \pm 0.069$, $\kappa = -0.2 \pm 0.10$, and $\xi = 12.83 \pm 0.069$, shown as the green long-dashed line in Figure~\ref{fig:AGN_frequency}. 
The subsequent CIC analysis is performed on this downsampled halo sample, employing the randomly selected cell placement scheme $(N_{\rm cell} = 10^9)$ described in Sec. \ref{sec:CIC}. 


\begin{figure}
    \centering
    \includegraphics[width=.45\linewidth]{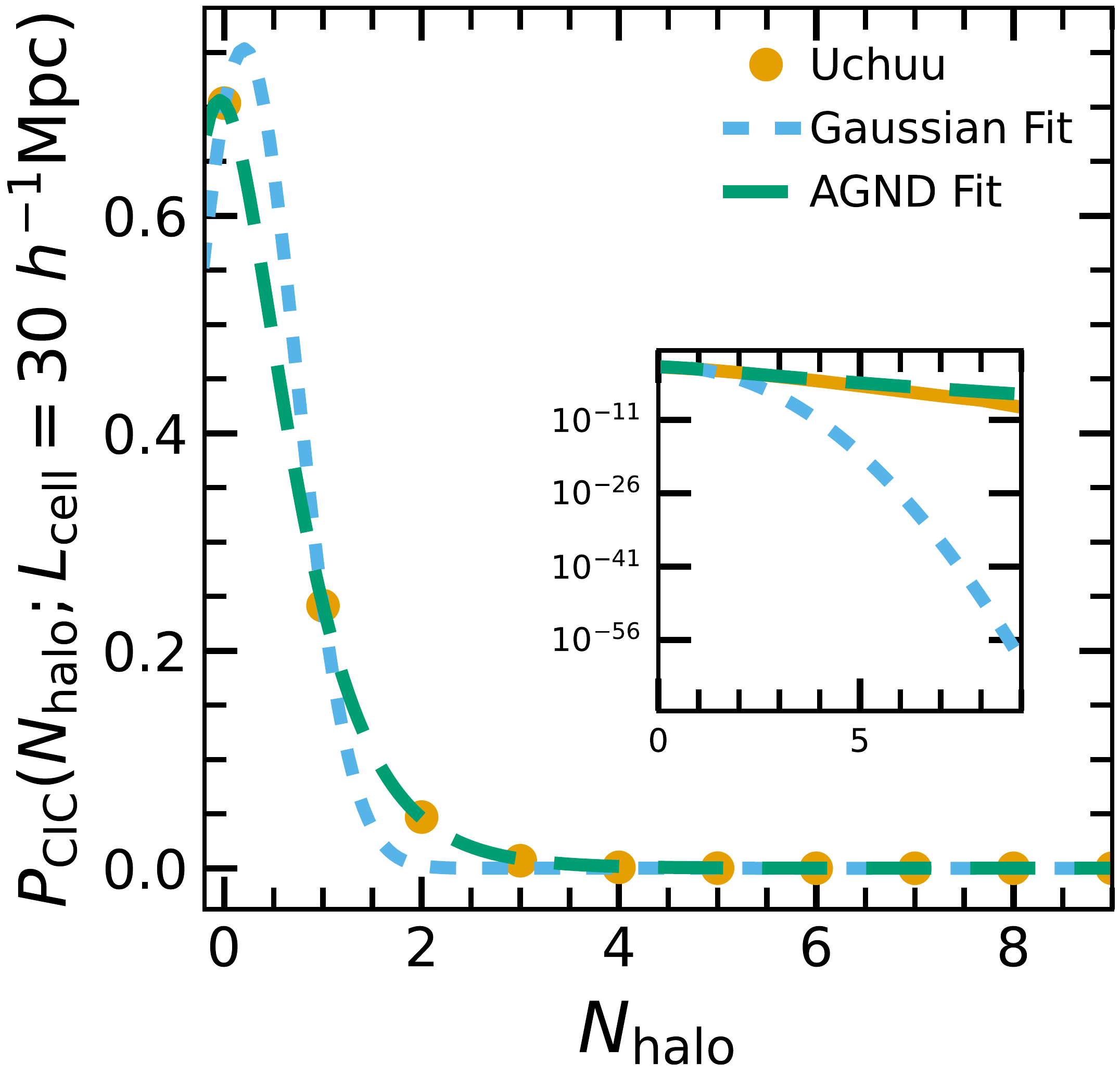}
    \caption{CIC probability distribution of quasar-host haloes for $L_{\rm cell} = 30 \ h^{-1} {\rm Mpc}$. The orange points show the CIC measured from the \textit{Uchuu} simulation \citep{ishiyama_uchuu_2021}. The blue dashed curve corresponds to a Gaussian fit, while the green long-dashed curve shows the AGND fit. The inset panel presents the same distributions on a semi-log scale, highlighting the deviations between the Gaussian model and the simulation results in the high-$\delta$ regime.}
    \label{fig:count-in-cells_haloes}
\end{figure}

Figure~\ref{fig:count-in-cells_haloes} presents the resulting CIC distributions for reduced haloes in the \textit{Uchuu} simulation. As with our \textit{CROCODILE} analysis, the fitting is performed using the non-linear least-square algorithm implemented in \texttt{scipy.optimize.curve\_fit} \citep{2020SciPy-NMeth}. Consistent with the results from our {\it CROCODILE} analysis, the Gaussian fit fails to reproduce the asymmetric and heavy-tailed shapes of the simulated distributions, whereas the AGND fit provides an excellent description across the full range of counts. This effect reproduces the unrealistically high-$N$ regime $(\delta_{\rm halo}^{\rm max} \sim 24 \sim 16.6 \sigma_{\rm Gauss})$ under the Gaussian assumption. 
The corresponding best-fit parameters are listed in Table~\ref{tab:Uchuu_fitting_parameters}. 

\par The result from this Gpc-scale analysis demonstrates that the non-Gaussian form of the CIC distribution persists in much larger volumes, confirming the robust existence of extreme quasar overdensities.

\begin{table}[htpb]
    \centering
    \caption{Best-fit parameters of the Gaussian and AGND model describing the halo CIC probability distributions in the \textit{Uchuu} simulation.}
    \label{tab:Uchuu_fitting_parameters}
    \begin{tabular}{ccc} \hline \hline
        Gaussian & $\mu$ & $\sigma$ \\ \hline 
        Case~C  &  $0.20 \pm 0.042$ & $0.53 \pm 0.028$ \\ \hline
    \end{tabular}
    \begin{tabular}{cccc} \hline \hline
        AGND & $\alpha$ & $\kappa$ & $\xi$ \\ \hline
        Case~C & $0.616 \pm 0.0014$& $0.416 \pm 0.0063$ & $0.190 \pm 0.0021$ \\ \hline
    \end{tabular} 
\end{table}

\section{The CIC results from different simulations}

To clarify the dependence of results on simulations, we applied the CIC analysis to the samples with wider mass ranges. For \texttt{BF50NEL} and \textit{Uchuu}, we selected haloes satisfying $\log_{10} (M_{h}/M_{\odot}) > 11.5$ and utilized the data at $z=2.2$, consistent with the redshift of our main analysis. Regarding Shin+2017, we utilized the data from \citet{shin_new_2017} at $z=0$ as a substitute, since the CIC distribution at $z\sim2$ was not reported in their study.
\par
\begin{figure}[h]
    \centering
    \includegraphics[width=0.5\linewidth]{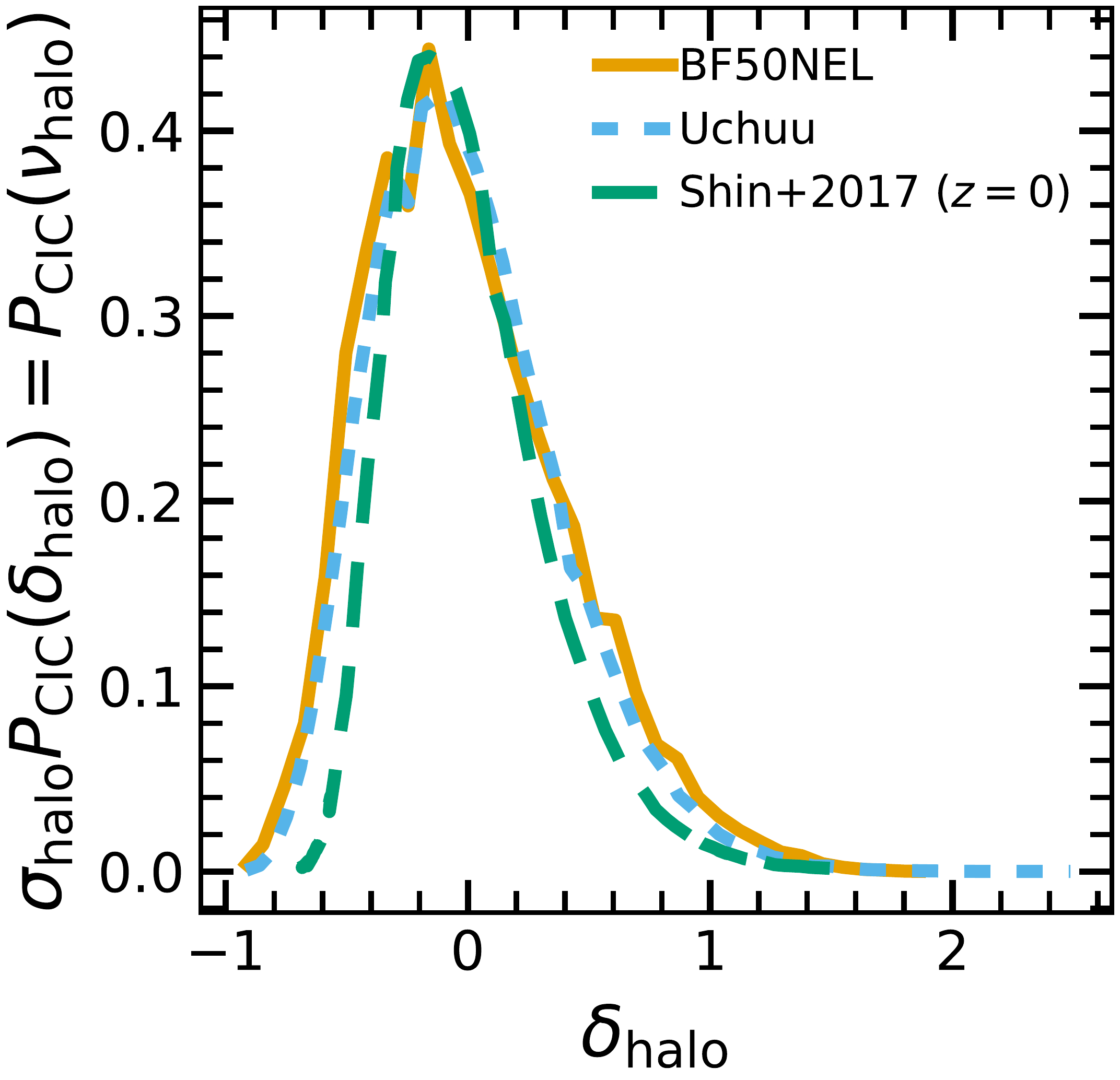}
    \caption{The CIC probability distribution for halos in three different simulations: \texttt{BF50NEL} with $\log_{10} M_h/M_\odot > 11.5$ (orange line), the \textit{Uchuu} simulation with $\log_{10} M_h/M_\odot > 11.5$ (skyblue dashed line), and $z=0$ data from \citet{shin_new_2017} (green long-dashed). For \texttt{BF50NEL} and the \textit{Uchuu}, we took cubic cells with  $L_{\rm cell}=30 \ h^{-1} {\rm Mpc}$, while spheres with a radius $R_{\rm th} = 25 \ h^{-1} {\rm Mpc}$ were used for \citet{shin_new_2017}.}
    \label{fig:count-in-cells_comparison}
\end{figure}

 Figure~\ref{fig:count-in-cells_comparison} shows the difference in the CIC for different simulations. The simulation of \citet{shin_new_2017} has a narrower distribution due to a higher resolution than ours and can model lower mass haloes. The CIC analysis of \textit{Uchuu} is based on the grid-based method, whereas random sampling ($N_{\rm cell} = 10^6$) was used for \texttt{BF50NEL}. The comparison demonstrates rough consistency in the derived quasar population in all simulations.

\bibliographystyle{aasjournal}
\bibliography{reference}
\end{CJK*}
\end{document}